%% file: main_arxiv.tex
\documentclass[12pt]{article}

\usepackage{arxiv}

\usepackage[utf8]{inputenc} % allow utf-8 input
\usepackage[T1]{fontenc}    % use 8-bit T1 fonts
\usepackage[hidelinks,hyperfootnotes=false]{hyperref}       % hyperlinks
\usepackage{url}            % simple URL typesetting
\usepackage{booktabs}       % professional-quality tables
\usepackage{amsfonts}       % blackboard math symbols
\usepackage{nicefrac}       % compact symbols for 1/2, etc.
\usepackage[nopatch=footnote]{microtype}      % microtypography
\usepackage{includes}
\renewcommand{\rev}[1]{#1}
\newcommand{\revii}[1]{#1}

\title{\titlename}

\author{
Molly Offer-Westort
\And
Drew Dimmery
}

\begin{document}
\maketitle

\begin{abstract}
\input{abstract.tex}

\end{abstract}

% keywords can be removed
%\keywords{First keyword \and Second keyword \and More}

\onehalfspacing
\input{body.tex}

\bibliographystyle{plainnat}  
\bibliography{references}

\clearpage
\input{appendix.tex}

\end{document}

%% file: abstract.tex
\rev{When interference is present, a unit's outcome depends on others' assignments and there is generally no single, design-free average treatment effect. 
We study settings where the decision problem is to choose among several alternative \emph{uniform} policies---e.g., rolling out one platform configuration to all users, or one policy to all constituents---so the estimand of interest is the Global Average Treatment Effect (GATE), the average outcome difference under two \emph{uniform} (global) policies.}
\rev{Under partial interference with clusters, cluster-level randomization identifies the GATE but can be statistically inefficient, while unit-level randomization can be highly precise yet biased for the GATE. 
We propose a continuum of implementable two-stage randomization designs that smoothly interpolate between these extremes by tuning within-cluster treatment correlation. For multi-arm experiments, we operationalize this continuum via a Dirichlet--multinomial design and give a pilot- and model-assisted procedure for selecting the design parameter using estimated finite-sample RMSE for the GATE. 
We also show that, even under a correctly specified linear interference model, difference-in-means estimators can have lower RMSE than least squares regression for GATE targets.}
\rev{Simulations and a large-scale Facebook video-player configuration experiment (43 million user sessions) illustrate the practical trade-off: intermediate designs can substantially reduce RMSE for estimating global rollout effects while remaining straightforward to deploy at scale.}

%% file: body.tex
\section{Introduction}
\rev{Many experiments are run to choose a single configuration that will be rolled out uniformly.
For example, in our motivating social media platform setting, each user session contains multiple videos; changing the video-player configuration for one video can affect the performance of other videos in the same session because they share a local resource (bandwidth). 
In a city governance setting, a policymaker may wish to choose among several alternative policies to implement at scale, and the delivery of one program to some subjects may affect responses of others.
The decision-maker's question is not ``what is the direct effect on a single video holding others fixed?'' but rather: \emph{which configuration would produce the best outcomes if applied to all videos for all users?} 
Or, in the policy setting, \emph{which policy would produce the best outcomes if applied to all constituents?}}

When the Stable Unit Treatment Value Assumption (SUTVA) is violated, a unit's potential outcomes are not only a function of the treatment assigned to them directly, but possibly also of the treatments assigned to other units, that is, there is ``interference'' \citep{cox1958}.
In such settings, the Average \rev{Treatment} Effect (ATE) is generally not uniquely defined, as a unit's treatment status may be associated with multiple potential outcomes under different treatment assignments of other units.
Researchers may instead turn to design-specific estimands, including a generalization of the ATE under interference, the Expected Average Treatment Effect (EATE) \citep{savje2021average}. 
\revii{The EATE averages direct-treatment contrasts over assignments of other units induced by the experimental design; we define it formally in Section~\ref{section:uniform-policies}. 
It arises as the probability limit of standard ATE estimators under broad forms of misspecified interference, which may not be the target for which the experiment was initially designed.}
\revii{Alternatively, the targets of interest may be decompositions of effects into direct and indirect effects as well as overall causal effects of the intervention \citep{halloran1995, hudgens2008, vanderweele2011effect}, or related targets such as effects on uniquely treated units and spillover effects on treated and untreated units as functions of treatment saturation \citep{baird2018optimal}.}
%
%Such interference has been studied at a peer-to-peer level, for example when interference occurs over a network \citep{aronow2016b,leung2016,bowers2013, bowers2015}, and when interference occurs at a group or cluster level and may be a function of treatment saturation \citep{miguel2004,hudgens2008,tchetgen2012causal,sinclair2012detecting}.

\rev{Our scope is experimental design and estimation for multi-arm experiments under partial interference.}
We consider average differences in unit-level outcomes under alternative uniform interventions, under which {all} units would be assigned the same treatment for each respective intervention.
We refer to this target as the Global Average Treatment Effect (GATE) \rev{\citep{chin2018regression, ugander2023randomized}}. 
Given the GATE as the estimand of interest, we identify a \revii{bias--variance trade-off} and show that root mean squared error (RMSE) under two-stage randomization procedures with intra-cluster correlation of treatment strictly between zero and one may dominate one-stage randomization procedures with randomization at either the unit or cluster level, under certain conditions.
We provide a principled way to make this choice by selecting from a continuum of possible intra-cluster correlations of treatments (see Figure~\ref{fig:homogeneous_target-design}). 
\rev{In our motivating application, a cluster is a user session and units are user--video impressions; partial interference means videos can compete within a session, but sessions do not affect one another.}
We then demonstrate performance of this approach through simulations.

%\rev{This paper makes three contributions. (i) We introduce a continuum of two-stage randomization designs for multi-arm experiments under partial interference, indexed by within-cluster treatment correlation, that interpolates between unit- and cluster-level randomization. (ii) For the GATE target, we characterize a bias--variance trade-off across this continuum and show that intermediate designs can minimize finite-sample RMSE. (iii) We show that, even when a linear interference model is correctly specified, a difference-in-means estimator can outperform least-squares regression in RMSE for GATE estimation, and we provide an implementation recipe (including a sequential ``pilot-then-tune'' workflow) suitable for large-scale online experimentation.}

\subsection{Related work and contributions}\label{sec:related-work}

\rev{A growing literature studies causal inference and experimental design under interference, where a unit's outcome depends on others' treatment assignments. Under \emph{partial interference} (\citealp{sobel2006randomized}), units are partitioned into disjoint clusters with interference allowed within but not across clusters, enabling design-based identification strategies; see, e.g., \citet{hudgens2008}, \citet{tchetgen2012causal}, and \citet{liu2014a}. More general approaches for arbitrary (known) exposure mappings are developed in \citet{aronow2017estimating}. Parallel work in online experimentation emphasizes graph clustering and cluster randomization as practical approximations to global policy effects in networked settings \citep{ugander2013graph,ugander2023randomized}, as well as variants such as ``hole-punching'' that trade bias and variance for network-exposure estimands \citep{eckles2017design}.}

\rev{Our inferential target is the \emph{Global Average Treatment Effect} (GATE), \revii{the average difference in outcomes under two uniform (global) assignment policies (formally defined in Section~\ref{section:uniform-policies}).} This target is natural when the decision problem is to choose a single configuration to roll out uniformly (as in our motivating platform application), and differs from estimands that decompose direct and spillover effects or vary treatment intensity within clusters.}

\rev{Our closest antecedent is \citet{baird2018optimal}, who formalize \emph{randomized saturation} designs for \emph{binary} treatment under partial interference and derive analytical standard errors and optimality results for estimating intention-to-treat and spillover estimands (and their slopes) as functions of treatment saturation. Their key message is a design trade-off: varying saturation identifies richer spillover structure but can reduce power for simpler pooled effects, especially when outcomes are strongly correlated within clusters. We build on the same partial-interference framework but address a different estimand and design problem that arises in multi-arm optimization and global policy evaluation.}

\rev{Our first contribution is to introduce a \emph{continuum} of two-stage randomization designs for \emph{multi-arm} experiments under partial interference that interpolates between cluster-uniform randomization (unbiased for GATE but potentially noisy) and unit randomization (highly precise but generally targeting a different estimand when interference is present). Concretely, we represent each cluster $j$ by a probability vector $\Pi_j$ over arms and assign units conditionally i.i.d.\ given $\Pi_j$; varying the dispersion of $\Pi_j$ tunes the induced within-cluster correlation of treatment assignments. This continuous tuning contrasts with designs that rely on a small, discrete set of saturations and is particularly convenient for large-scale implementation where one may want to smoothly trade off bias and variance.}

\rev{Our second contribution is to frame design selection for global rollout evaluation as an explicit \emph{bias--variance trade-off for the GATE}. Under partial interference, non-uniform assignments generally shift the conditional exposure environment of treated and control units away from the uniform-policy counterfactuals that define the GATE. At the same time, reducing within-cluster correlation of treatment can dramatically reduce variance by increasing the effective number of independent observations. We show how intermediate designs can minimize finite-sample RMSE for GATE estimation across arms, providing a principled rationale for designs that are neither fully cluster- nor fully unit-randomized.}

\rev{Our third contribution concerns estimation. Using a linear interference model that generalizes regression analyses used for saturation designs to a multi-arm setting, we connect model parameters directly to uniform-policy counterfactual means and then to the GATE. 
We then show that, for the GATE target and finite samples, a difference-in-means estimator can outperform least squares in RMSE even when the regression model is correctly specified---a result with direct practical implications for experimentation workflows where simple estimators are favored for robustness and operational simplicity.}

\revii{The linear model is a transparent working model that facilitates comparison of estimators for the same target. 
Correct specification is not required to define the design family or to identify the GATE under cluster-uniform randomization. 
It \emph{does}, however, facilitate the particular plug-in bias and variance calculations used to tune an intermediate design, which motivates cluster-uniform validation.}

\rev{Together, these contributions reposition two-stage designs as tools not only for identifying and measuring spillovers, but also for \emph{efficiently learning global rollout effects} in multi-arm experiments where the ultimate objective is policy selection under partial interference.}

\begin{figure}[!ht]
\centering
\includegraphics[width=.6\textwidth]{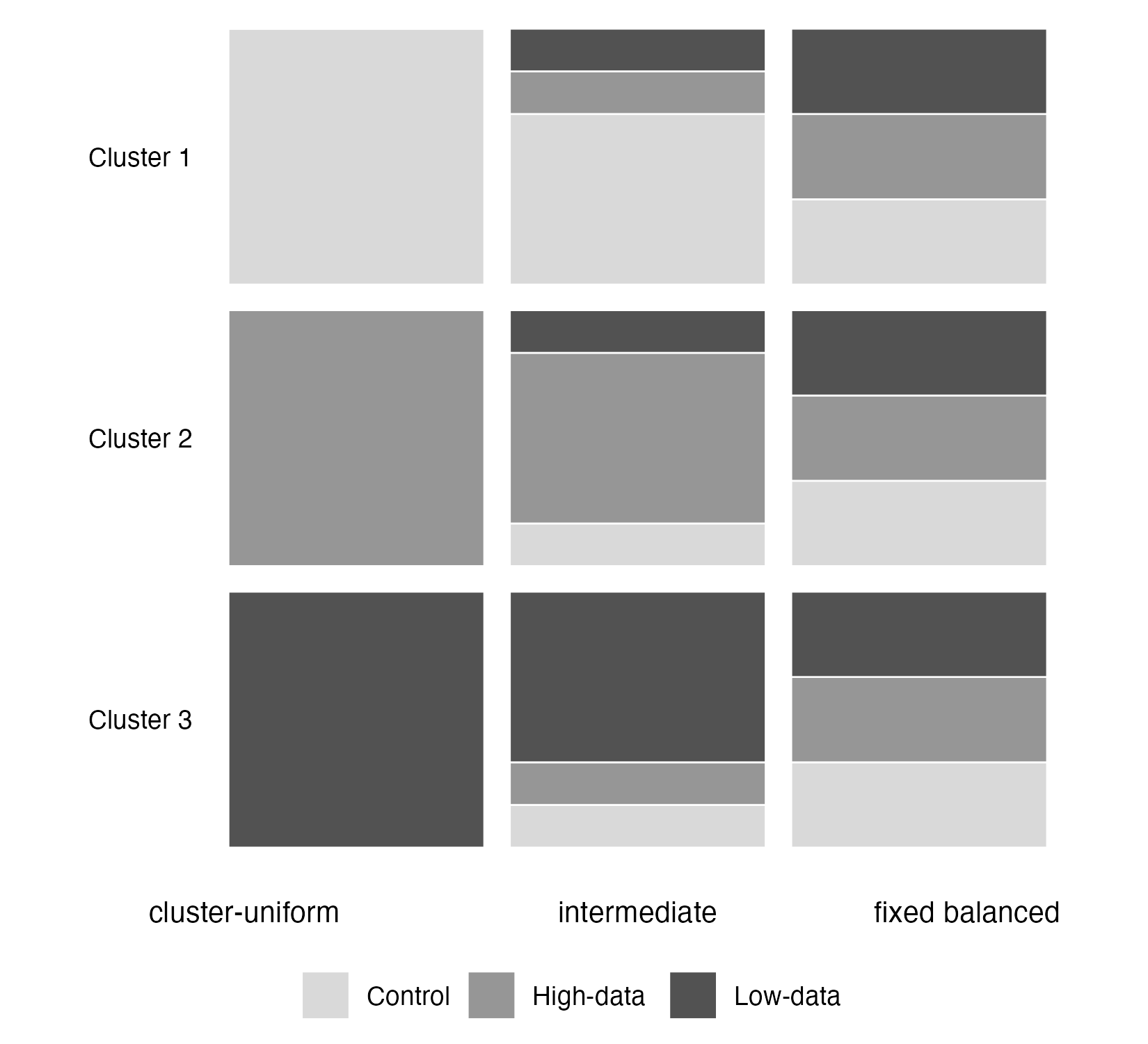}
\caption{\textbf{Assignment regimes ranging from cluster-uniform assignment to extensive within-cluster mixing.}}
\label{fig:homogeneous_target-design}
\end{figure}

\clearpage

\begin{example*}
\noindent\textbf{Toy setup.}
\rev{There are $J=3$ user sessions (clusters), each containing $N=6$ videos (units), and each video is assigned to one of three arms $A_{j,i}\in\{0,1,2\}$ \revii{(control, high-data, low-data)}.
The decision problem is to choose a \emph{single} configuration to roll out uniformly, so the estimand for arm $m\in\{1,2\}$ is the uniform-policy contrast $\Psi_m^{\mathrm{GATE}}=\E[Y_{j,i}(\bar\m_j)-Y_{j,i}(\bar{\bm{0}}_j)]$.
\revii{For each $m$, the toy difference-in-means estimator is $\widehat\Delta_m=\overline Y_{A=m}-\overline Y_{A=0}$, where $\overline Y_{A=a}$ averages observed outcomes among videos assigned to arm $a$. Write $\A_j=(A_{j,1},\ldots,A_{j,N})^\top$ for the assignment vector in session $j$. Let $n_{j\ell}=\sum_{r=1}^N\mathbf{1}\{A_{j,r}=\ell\}$ be the number of videos assigned to arm $\ell$ in session $j$, and let $\mu_j$ be that session's fixed baseline mean. Conditional on the assignments within session $j$, suppose that
\begin{align*}
\E[Y_{j,i}\mid A_{j,i}=0,\A_j]&=\mu_j,\\
\E[Y_{j,i}\mid A_{j,i}=1,\A_j]&=\mu_j+0.5-0.05\,n_{j2},\\
\E[Y_{j,i}\mid A_{j,i}=2,\A_j]&=\mu_j+0.75+0.05\,n_{j1}.
\end{align*}
For this toy calculation, outcomes are taken to equal these conditional means. The fixed baseline cancels when comparing uniform policies, so $\Psi_1^{\mathrm{GATE}}=0.5$ and $\Psi_2^{\mathrm{GATE}}=0.75$. For each regime, the three composition patterns specified for that regime are assigned to the three sessions at random; call this randomization $D$. Thus the only randomness is which pattern is assigned to which session, holding $(\mu_1,\mu_2,\mu_3)$ fixed. Write $\bar\mu=J^{-1}\sum_{j=1}^J\mu_j$ and $S_\mu^2=(J-1)^{-1}\sum_{j=1}^J(\mu_j-\bar\mu)^2$.}
See Appendix~\ref{app:toy} for the full setup and calculations.}

\rev{\emph{Three assignment regimes (Figure~\ref{fig:homogeneous_target-design}).}
\begin{enumerate}\setlength\itemsep{2pt}
\item \revii{\textbf{Cluster-uniform}: sessions are assigned a single arm, so $\E_D[\widehat{\Delta}_1]=0.5$ and $\E_D[\widehat{\Delta}_2]=0.75$. Randomly assigning these patterns across heterogeneous fixed sessions contributes $2S_\mu^2$ to $\operatorname{Var}_D(\widehat\Delta_m)$.}
\item \revii{\textbf{Fixed balanced complete randomization}: each session assigns exactly two videos to each arm, yielding biased rollout estimates ($\widehat{\Delta}_1=0.4$, $\widehat{\Delta}_2=0.85$). Equal arm weights within every session make the fixed baselines cancel exactly, so their variance contribution is zero.}
%Intuitively, these contrasts correspond to outcomes under a \emph{mixed} session composition (here, approximately $p_j[1]=p_j[2]=1/3$), rather than under the homogeneous policies $\bar 1$ and $\bar 2$ that define the GATE.
\item \revii{\textbf{Intermediate two-stage}: mostly uniform sessions with some mixing (e.g., 4 control, 1 high-data, 1 low-data) reduce bias relative to full mixing while retaining some variance reduction: $\E_D[\widehat{\Delta}_1]=0.425$ and $\E_D[\widehat{\Delta}_2]=0.825$, while the fixed-baseline contribution to $\operatorname{Var}_D(\widehat\Delta_m)$ is $S_\mu^2/2$.}
\end{enumerate}}
\end{example*}

\section{Motivating application}
Our motivating application is an analytic setting at Facebook, where experiments are run to tune the performance of the video player.
The treatments represent combinations of engineering parameters determining the behavior of video playback on the platform; while substance of these treatments is not of general substantive interest (e.g., changing how much data is pre-buffered on a video) it is vital to improving the overall experience on Facebook.
When a user scrolls past or interacts with videos on their Facebook timeline, multiple videos may load and play in quick succession.
There is a fixed quantity of resources at the cluster level (a user's bandwidth), and individual videos compete for this resource in response to changes in configuration;
these parameters ensure that videos load and play with minimal error.
The end goal of this experimentation is to launch a single configuration to all users and all videos which will provide the best user experience, as operationalized by internal video playback metrics.
That is, we ask, ``During experimentation, how should videos for each user be distributed among the various configurations to ensure we learn best about what would happen if we rolled out this change to everyone?''

The experiments we analyze here were conducted on the platform in June and July of 2018, across 43 million unique sessions where a user logged onto the platform.
On average, each session included 9.25 videos.
Parameters were tuned according to the typical internal process for testing configuration changes for the Facebook web video player; as such, they would typically be benign and invisible to the user.
Since \revii{these experiments were} deployed on Facebook in a real-world setting, the subject pool is sampled from the diverse Facebook user base: different users have different internet connections and different content available to view through the Facebook feed.

	The assumption of partial interference is well-motivated here: in this setting, there is meaningful interference, but it is constrained within clusters: making one video on a user's feed play more seamlessly may hurt the performance on another video, but it is implausible for a video's performance on one person's timeline to affect the performance of a video \rev{for another user}.
	The goal of a global assignment policy is also a natural objective, as the platform runs experiments with the goal of determining a uniform policy rule prospectively (in practice, this decision rule may be stratified by, for example, connection class or device type, but uniform within user classes).
	\rev{More global interactions (e.g., system-wide congestion or ranking feedback) are possible in platform settings; in our application the experiment is balanced and small relative to total traffic and assignments are made at the session level, so any system-wide effects are plausibly common across arms. We therefore treat sessions as independent clusters.}

	\rev{The Facebook video experiment can also be viewed through the lens of experimentation in two-sided platforms and marketplaces \citep{johari2022experimental,bajari2023experimental}, where users and items represent two sides of the market and interventions can in principle affect both sides.} \revii{However, we adopt the partial-interference framework here because the natural cluster definition (user sessions) aligns with the interference structure.} \rev{This perspective would be more relevant if we were concerned with interference through supply-side responses (e.g., content supply or creator incentives), which we explicitly exclude from scope.}

	While our motivating application is online experimentation, we argue that this structure is also relevant in social, public health, and policy \revii{settings---such} as when clusters represent schools, communities, or households, where interference will be more meaningful within than across clusters and where decision-makers aim to implement scalable, population-wide programs.
	Clusters share a local constraint (bandwidth, attention, capacity, budget), so assigning a more \emph{intensive} condition to one unit can change outcomes for other units in the same cluster.

\section{Notation and setting}\label{section:setting}

\rev{Consider an experiment over a set of clusters, indexed by $j = 1, \dots, J$, each composed of $N$ individual units.}\footnote{For simplicity of exposition, we fix all $N_j$ as equal, i.e.,  $N_j = N$ for all $j$ in $1, \dots, J$. \rev{The plug-in RMSE calculations in Section~\ref{sec:plugin-rmse} show how we incorporate unequal cluster sizes in applications.}} %TODO {check if this is necessary for any results}
We assume i.i.d. draws from an infinite \revii{superpopulation} of clusters.
Let $\A_j=(A_{j,1}, \dots, A_{j,N} )^\top$ be the $N$-dimensional column vector of assignments in cluster $j$, where $A_{j,i} \in \{0, \dots, M \}$ determines which of $M+1$ treatments unit $i$ receives, with level zero being the control or status quo.
Outcomes for units in cluster $j$ are defined as $\Y_j = (Y_{j,1}, \dots, Y_{j,N})^\top$.
The probability distribution of the random vector $\A_j$ is completely determined by the experimental design set by the researcher; we discuss randomization procedures below. \rev{Let $\A=(\A_1^\top,\ldots,\A_J^\top)^\top$ denote the full $JN$-dimensional column vector across clusters. \revii{Equivalently, one may arrange the same assignments as a $J\times N$ matrix with row $j$ equal to $\A_j^\top$; throughout we use the stacked-vector convention.}}

\subsection{Partial interference and potential outcomes}
\begin{assumption}\label{ass:partial}
\revii{Let $\aB$ and $\aB'$ be fixed realizations of the full assignment vector $\A$. Under partial interference \citep{sobel2006randomized}, a unit's potential outcome may depend on the full assignment vector within its cluster but not on assignments in other clusters:}
\begin{align}
\revii{Y_{j,i}(\aB)=Y_{j,i}(\aB')
\quad\text{for all }i,j,\aB,\aB'\text{ such that }\aB_j=\aB'_j.}
\end{align}
\revii{We may therefore write the potential outcome as $Y_{j,i}(\aB_j)=Y_{j,i}(a_{j,i},\aB_{j,-i})$, where $\aB_{j,-i}$ removes unit $i$ from $\aB_j$.}
\end{assumption}

\subsection{\rev{Uniform (global) policies}}\label{section:uniform-policies}
\revii{For comparison with our target, we first formally define the EATE now that the assignment notation is in place. Under a specified assignment design $D$, let $\A_{j,-i}$ denote the assignments of the other units in cluster $j$, and let $\E_{j,i}$ denote the unit-weighted expectation over sampled units. The EATE comparing direct assignments $m$ and $0$ is
\begin{align}
\Psi^{\mathrm{EATE}}_D(m,0)
=\E_{j,i}\E_{\A_{j,-i}\sim D}
\left[Y_{j,i}(m,\A_{j,-i})-Y_{j,i}(0,\A_{j,-i})\right].
\label{eq:eate}
\end{align}
Thus, the EATE marginalizes a unit's direct-treatment contrast over the assignments of other units induced by $D$ \citep{savje2021average}.}

We are interested, however, in comparing counterfactual outcomes under two specific \rev{uniform policies}, each of which assigns the same treatment level to all units.
Consider for example a setting where we want to change from the status quo, under which all units receive the same treatment, to some alternative policy, which will also be implemented \rev{uniformly} across all units.
In this setting, we do not need to take the step of marginalizing over all treatment vectors, as each unit has a single potential outcome under each setting.
\revii{Let $\bar\m_j=(m,\ldots,m)^\top$ and $\bar{\bm{0}}_j=(0,\ldots,0)^\top$ denote the uniform assignment vectors in cluster $j$; we omit the cluster subscript when referring to the corresponding global policies.}

We term the \rev{uniform-assignment} unit-level treatment effect as the unit-level treatment effect of being assigned $m$ when all other units receive assignment $m$, as compared to being assigned zero when all other units receive assignment zero.\footnote{This definition is flexible to fixing any treatment level as the control, and can be used for comparison of arbitrary treatment levels. }
As shorthand, we may write,
 \begin{align}
 \Psi_{j,i}(\bar \m_j,\bar{\bm{0}}_j) = Y_{j,i}(\bar \m_j ) - Y_{j,i}(\bar{\bm{0}}_j).
 \end{align}

Taking the expectation over the distribution of units, our inferential target is the Global Average Treatment Effect (GATE), which is defined as,
\begin{align}
\Psi^{\mathrm{GATE}}(\bar \m, \bar{\bm{0}}) & =\E\left[ Y_{j,i}(\bar \m_j) -  Y_{j,i}( \bar{\bm{0}}_j) \right] .\footnotemark \label{eq:gate}
%& = \E[ \Psi_{j,i}(\aB, \aB')].
\end{align}
\footnotetext{\revii{The expectation is unit-weighted. With equal cluster sizes, it also equals the expectation of the within-cluster average over clusters; with unequal sizes, the corresponding cluster average must be weighted by cluster size.}}
\revii{For brevity, write $\Psi_m^{\mathrm{GATE}}:=\Psi^{\mathrm{GATE}}(\bar\m,\bar{\bm{0}})$.}
This estimand has been considered by others, including \cite{baird2018optimal}, \cite{basse2018limitations}, and \cite{chin2018regression}. % \begin{align}
% \Psi^{GATE} = \sum_{j = 1}^J\sum_{i = 1}^{N}\Psi_{j,i}(a,a')
% \end{align}
%[[more fully explain how this relates to H\&H; I think it's average total causal effect for specific conditions; this gets around the Hudgens \& Halloran problem of marginalizing over different distributions when defining some of their unit-level treatment effects, which makes for a not well-defined estimand. And it means we don't have to marginalize in the EATE, but we're able to compare two different designs, albeit very specific ones, but of broad interest. ]]
This estimand is well-defined under arbitrary forms of interference, but without additional assumptions for example on the nature of interference, the effect is generally not identifiable.
Under no interference, the treatment assignment vector for other units does not affect the potential outcome, and the GATE, the EATE, and the ATE will all be equivalent.

\section{Identification and estimation}
Having assumed partial interference, however, it is the case that a unit's potential outcomes are only defined by the assignments of treatment within their cluster, regardless of treatment assignments in other clusters.
We can thus use cluster-uniform random assignment, under which every unit in a cluster receives the same arm and clusters are randomized between $\bar \m_j$ and $\bar{\bm{0}}_j$, to identify the GATE.

\subsection{Difference-in-means}
We can use the difference in sample means for estimation of the GATE.
\begin{align}
\widehat\Psi_m^{\mathrm{DM}}& =  \frac{  \sum_{j = 1}^J \sum_{i = 1}^{N} Y_{j,i} \mathbf{1}\{ A_{j,i} = m \}}{  \sum_{j = 1}^J \sum_{i = 1}^N \mathbf{1}\{ A_{j,i} = m\}} -   \frac{  \sum_{j = 1}^J \sum_{i = 1}^{N}  Y_{j,i} \mathbf{1}\{ A_{j,i} = 0 \}}{ \sum_{j = 1}^J  \rev{\sum_{i = 1}^N} \mathbf{1}\{ A_{j,i} = 0\}} .
\label{eq:DM}
\end{align}

\rev{
\begin{proposition}[Unbiasedness of difference-in-means for the GATE]\label{prop:dm_unbiased}
\revii{Under the standing equal-cluster-size condition $N_j\equiv N$, Assumption~\ref{ass:partial}, and complete cluster randomization that assigns fixed positive numbers of clusters to $\bar \m_j$ and $\bar{\bm{0}}_j$, independently of potential outcomes, the difference-in-means estimator $\widehat\Psi_m^{\mathrm{DM}}$ in \eqref{eq:DM} is unbiased for the GATE $\Psi_m^{\mathrm{GATE}}$ in \eqref{eq:gate}.}\footnote{\revii{Proposition~\ref{prop:dm_unbiased} establishes exact unbiasedness under equal cluster sizes. With unequal cluster sizes, Equation~\eqref{eq:DM} is a ratio estimator and is generally not exactly design-unbiased, although it is consistent under standard regularity conditions as the number of independent clusters grows.}}
\end{proposition}
\noindent\emph{Proof.} See Appendix~\ref{app:proof-dm-unbiased}.
}

If we wanted to identify the EATE \citep{savje2021average}, or other design-specific estimands \citep{aronow2025nurva}, one general approach would be to use the Horvitz--Thompson estimator \citep{horvitz1952}, weighting by the inverse probability of the realized assignment.
\cite{aronow2025nurva} show that the \revii{Horvitz--Thompson} estimator can be seen as a generalization of the sample mean.

\subsection{Linear model}\label{section:lim}
Linear-in-means models have been proposed in settings with social interaction where outcomes for an individual are a function of outcomes for a related group \citep{manski1993identification,bramoulle2009identification}.
In such models, a unit's outcome is a linear function of the mean characteristics of their group, and possibly a set of the unit's own characteristics.
\cite{chin2018regression} has developed such models to account for interference in Bernoulli randomized trials with covariates that are arbitrary functions of the treatment assignment vector; %Linear-in-means models \citep{manski1993identification} were originally developed to describe settings with social interaction and endogenous effects. The model has been expanded and built on by \cite{bramoulle2009identification}
%\cite{chin2018regression}
\cite{baird2018optimal} consider regression models for randomized treatment saturation designs with binary treatments and a fixed set of saturations determined by the researcher.

Here, we propose a model where individual response is a function of direct treatment as well as treatment assigned to the group as the underlying \revii{data-generating} process; generalizations are possible, but the simple model will allow us to more clearly demonstrate bias and variance trade-offs.
The linear model need not be true for our recommended randomization procedures; our goal is to show that \textit{even when a regression model is correctly specified}, the least squares regression estimator may be an inferior estimator to simple difference in means estimators for the target estimand.

\rev{For the purposes of this illustrative estimator, we} follow in the spirit of \cite{baird2018optimal} and  \cite{chin2018regression}, with generalization to the clustered, multiple treatment setting.
The model proposed by \cite{baird2018optimal} includes an intercept and separate indicators for directly treated units and untreated units in treated clusters at each assigned saturation level. %[[just include the model??]]
They estimate slopes separately for treated and untreated units in treated clusters as the difference in coefficients on indicators at two different treatment saturation levels, divided by the difference in saturation levels.
To allow for more flexible design selection in the multiple treatment setting, we will include slope coefficients directly in the model.

\begin{assumption}[\rev{Outcome dependence on exposure}]\label{ass:proportional}
\rev{
Expected potential outcomes depend \emph{only} on the unit's direct treatment $a_{j,i}$ and the within-cluster treatment shares
\[
p_{j[m]}(\aB_j) = \frac{1}{N}\sum_{i'=1}^N \mathbf{1}\{a_{j,i'}=m\},\qquad m\in\{0,\ldots,M\}.
\]
That is, there exists a function $g$ such that for all treatment vectors $\aB_j$,
\begin{align}
\E\!\left[ Y_{j,i}(\aB_j )\right] = g\!\left(a_{j,i};\  p_{j[0]}(\aB_j), \dots, p_{j[M]}(\aB_j)  \right).
\end{align}
Equivalently, this defines an exposure mapping in the sense of \citet{manski2013identification} and \citet{aronow2017estimating}: if two assignment vectors $\aB_j$ and $\aB'_j$ satisfy $a_{j,i}=a'_{j,i}$ and $p_{j[m]}(\aB_j)=p_{j[m]}(\aB'_j)$ for all $m$, then $\E[Y_{j,i}(\aB_j)]=\E[Y_{j,i}(\aB'_j)]$.
	\footnote{To avoid indexing by units as well, and because slope effects are modeled separately for each direct treatment condition, we include all units in the share so that each unit in the same cluster shares the same values of $p_{j[m]}$ for all $m$ in $0, \dots, M$.}
}
\end{assumption}

Note that the respective treatment proportions within a cluster $j$ must sum to one:
\begin{align}
\sum_m p_{j[m]}(\A_j) = 1.
\end{align}

Following \cite{baird2018optimal}, this is a modification of the stratified interference assumption \citep{hudgens2008}, allowing realized potential outcomes to depend on the units that receive each treatment assignment, but constraining expected potential outcomes to simplify representation of standard errors without network knowledge.\footnote{With finite cluster size, the treatment-share exposure takes values on a discrete support, so the representation does not require continuity of the exposure-response function between attainable shares.}

\begin{assumption}\label{ass:homoskedasticity}
\revii{Given the i.i.d. sampling of clusters assumed above, we follow \cite{baird2018optimal} in imposing homoskedasticity over potential outcomes and an exchangeable within-cluster covariance structure such that, for all treatment vectors,}
\begin{align}
\Var[Y_{j,i}(\aB_j) ] &= \sigma^2 + \tau^2,\\
\Cov[Y_{j,i}(\aB_j), Y_{j,i'}(\aB_j) ] &= \tau^2 \textrm{ for }i \neq i'.
\end{align}
Intra-cluster correlation of errors is then
\begin{align}
\rho_u = \frac{\tau^2}{\tau^2 + \sigma^2}.
\end{align}
\end{assumption}
\revii{For model-based estimation and plug-in tuning, we further adopt the following linear working model for $g$:}
\begin{align}
Y_{j,i}
=\sum_{m=0}^M \beta_m \mathbf{1}\{A_{j,i}=m\}
+ c \sum_{m=0}^M \sum_{\ell=1}^M \delta_{m,\ell}\,\rev{p_{j[\ell]}(\A_j)}\,\mathbf{1}\{A_{j,i}=m\}
+\varepsilon_{j,i}.
\label{eq:linmod}
\end{align}
We interpret $c$ as an interference-magnitude scalar that scales the exposure-response terms; since only the products $c\,\delta_{m,\ell}$ are identified, we absorb $c$ into $\delta$ (equivalently, set $c=1$) except when varying interference strength in simulations (Section~\ref{section:sim}).
The error term is the difference between the realized unit potential outcome as a function of the treatment vector, and the unit expected potential outcome for that treatment vector. %, expressed as $\E\left[ Y_{j,i}(a_{j,i}; \  p_{j[0]}(\A_j), \dots, p_{j[M]}(\A_j)  )\right]$.
Under the assumptions imposed above, $\E[\varepsilon_{j,i}|\A_j ] = 0$ \citep[Lemma 1]{baird2018optimal}. \revii{When the working model is correctly specified and the design matrix has full column rank, OLS is unbiased for the model coefficients and hence the plug-in contrast defined below is unbiased for the model-implied GATE.}
% (See \ref{homog_appendix:unbiased}). TKTK

\rev{Under a uniform (global) policy $m$, every unit in cluster $j$ has $A_{j,i}=m$ and $p_{j[m]}(\bar \m_j)=1$ with $p_{j[\ell]}(\bar \m_j)=0$ for $\ell\neq m$. Plugging into Equation~\ref{eq:linmod} (absorbing $c$ into $\delta$) yields $\E[Y_{j,i}(\bar \m_j)]=\beta_m+\delta_{m,m}$, while under uniform control $\E[Y_{j,i}(\bar{\bm{0}}_j)]=\beta_0$. Therefore the GATE for policy $m$ is}
\begin{align}
\rev{\Psi_m^{\mathrm{GATE}} = (\beta_m+\delta_{m,m})-\beta_0.}
\end{align}
\rev{Under Assumptions~\ref{ass:proportional}--\ref{ass:homoskedasticity}, a natural model-based estimator of this GATE contrast is the plug-in estimator}
\begin{align}
\rev{\widehat\Psi_m^{\mathrm{LM}} = (\widehat{\beta}_{m}+\widehat{\delta}_{m,m})-\widehat{\beta}_0,}
\end{align}
\rev{where $(\widehat{\beta},\widehat{\delta})$ are obtained by OLS fit of Equation~\ref{eq:linmod}.} \revii{Under correct specification, positivity, and full column rank, this estimator targets the same uniform-policy contrast $\Psi_m^{\mathrm{GATE}}$, but its finite-sample RMSE can be larger than difference-in-means because it estimates additional nuisance parameters governing interference.}
\rev{The remaining $\delta_{m,\ell}$ terms quantify how non-uniform within-cluster mixes shift average outcomes away from the uniform-policy counterfactuals.}

%And, also as with \cite{chin2018regression}, $\beta_m$ and $\delta_{m, m}$ and $\beta_0$ are only nuisance parameters, and we are primarily interested in $\E[Y_{j,i}(\bar \m_j)] - \E[Y_{j,i}(\bar{\bm{0}}_j)]$.

%[[should this be in terms of rho?]]
%That is, $\beta_m$ terms represent the expected potential outcome for being uniquely treated in a cluster with all other units taking on the control status, and each of the $\delta_{m, \ell}$ terms allow different slopes
%\begin{align}
%Y_{j,i} = \sum_{m=0}^M \beta_m \{ a_{j,i} = m\} + \sum_{m=0}^M \sum_{\ell=1}^M \delta_{m, \ell} p_{j,[\ell]} \{ a_{j,i} = m\}  + \varepsilon_{j,i }.
%\end{align}
%%[[should this be in terms of rho?]]
%That is, $\beta_m$ terms represent the expected potential outcome for being uniquely treated in a cluster with all other units taking on the control status, and each of the $\delta_{m, \ell}$ terms allow different slopes
%\begin{align}
%Y_{j,i} & = \beta_0 \{a_{j,i} = 0 \} + \delta_{0, 1}  \{ a_{j,i} = 0\} p_{j,1} + \delta_{0, 2}  \{ a_{j,i} = 0\} p_{j,2} \\
%& + \beta_1 \{a_{j,i} = 1 \}+ \delta_{1, 1}  \{ a_{j,i} = 1\}p_{j,1} + \delta_{1, 2} \{ a_{j,i} = 1\} p_{j,2}+ \\
% & \beta_1 \{a_{j,i} = 1 \} + \delta_{1, 1}  \{ a_{j,i} = 1\} p_{j,1}+ \delta_{1, 2} p_{j,2}\{ a_{j,i} = 1\} + \varepsilon_{j,i}
%\end{align}

%That is, we allow for the underlying potential outcomes to vary as a function of the entire treatment vector within a cluster, but we are only interested in estimating average outcomes as a function of direct treatment and the proportion of other units in the cluster that share the same treatment.

\revii{When randomization is not uniform at the cluster level, the model demonstrates why the difference-in-means estimator can be biased for the GATE in this setting;} Section~\ref{sec:bias} makes this bias explicit.

In this model, we have assumed that interference is a linear function of proportion of each type of treatment within a cluster; this can be broadened to consider the type of general ``interference control variables,'' discussed in \cite{chin2018regression}, Section 3, which are a function of the vector $\A_{j,-i}$.
We could define such a deterministic function, following \cite{chin2018regression}, as  $X_{j,i}(\A_{j,-i})$, to account for non-linearities in interference, endogenous effects, or to encode further information about the network structure or other ways that interference is channeled.
This requires independence of potential outcomes with treatment assignment, and equivalent exogeneity assumptions on the errors as those imposed by our assumptions above. %TODO {Note above that we're going to allow for further generalizations?}

\section{\rev{Randomization procedures and RMSE-guided tuning}}\label{sec:design}

\rev{
\revii{For each candidate policy $m\in\{1,\ldots,M\}$, let $\Psi_m^{\mathrm{GATE}}$ denote the uniform-policy contrast with control defined in Equation~\eqref{eq:gate}; with unequal cluster sizes, this is the corresponding unit-weighted contrast for the planned size profile.} Let $D$ denote an experimental design, i.e., a distribution over the stacked assignment vector $\A$, and \revii{let $\Y^{\mathrm{obs}}$ denote the corresponding stacked vector of observed outcomes.} Let $\widehat\Psi_m=\widehat\Psi_m(\A,\Y^{\mathrm{obs}})$ denote an estimator of $\Psi_m^{\mathrm{GATE}}$ (e.g., the difference-in-means estimator in \eqref{eq:DM}).
\revii{We evaluate designs by their repeated-sampling root mean squared error, conditioning throughout on positive realized counts for every arm entering the design objective. Operationally, an experiment-wide assignment with an empty required arm is rejected and redrawn in full:}
\[
\mathrm{RMSE}_D(\widehat\Psi_m)
=\revii{\left\{\E_{\mathcal S,D}\!\left[\big(\widehat\Psi_m-\Psi_m^{\mathrm{GATE}}\big)^2\right]\right\}^{1/2}},
\]
\revii{Here $\mathcal S$ denotes the sample of cluster potential-outcome schedules, drawn independently conditional on the fixed cluster-size profile, and the expectation averages over cluster sampling and the retained random assignments from $D$ after applying this redraw rule. We also report the design-specific probability that the initial assignment has positive counts in every required arm and do not select candidate designs for which an invalid initial assignment occurs with non-negligible probability. Any model-based estimator used in this criterion must specify how a non-estimable target contrast would be handled.}
All comparisons hold fixed the experimental budget, taken here to be the number of clusters $J$ and the cluster sizes $\{N_j\}_{j=1}^J$ (or $N$ under the equal-size simplification in Section~\ref{section:setting}), i.e., the total number of observed unit outcomes $\sum_{j=1}^J N_j$ is fixed.
In what follows we restrict attention to the two-stage probability-vector designs described in Section~\ref{sec:two-stage}, indexed by a single scalar tuning parameter $\bar\alpha$ (equivalently the within-cluster treatment correlation $\rho_T$). Accordingly, $\mathcal{D}=\{D(\bar\alpha):\bar\alpha\in\mathcal{A}\}$ for a prespecified grid or interval $\mathcal{A}$.
Our design objective is to choose $D$ to minimize average RMSE across the $M$ policy contrasts, for a fixed experimental budget:
\begin{equation}\label{eq:design-objective}
D^\star \in \arg\min_{D\in\mathcal{D}}\ \frac{1}{M}\sum_{m=1}^M \mathrm{RMSE}_D(\widehat\Psi_m).
\end{equation}

We focus on control-referenced contrasts because these are the primary quantities used to rank and select candidate policies relative to a status quo in multi-arm experiments. 
If a setting requires a different loss, the framework above extends by replacing \eqref{eq:design-objective} with (i) a weighted average $\sum_{m=1}^M \lambda_m\,\mathrm{RMSE}_D(\widehat\Psi_m)$, where $\lambda_m\geq 0$ and $\sum_{m=1}^M\lambda_m=1$, to reflect unequal importance across arms, (ii) a worst-case criterion $\max_{m\le M}\mathrm{RMSE}_D(\widehat\Psi_m)$, or (iii) an objective over all pairwise contrasts $\Psi^{\mathrm{GATE}}(\bar \m,\bar \m')=\Psi^{\mathrm{GATE}}(\bar \m,\bar{\bm{0}})-\Psi^{\mathrm{GATE}}(\bar \m',\bar{\bm{0}})$.
We also restrict attention below to designs that satisfy positivity and (for exposition) balanced marginal assignment across arms; \revii{Section~\ref{sec:dirichlet} shows how to accommodate alternative positive marginal allocation targets.}

Our ultimate design objective, given a fixed cluster definition and a GATE target, is to choose the within-cluster treatment correlation (our $\bar\alpha$ or $\rho_T$) to minimize an estimable approximation to finite-sample RMSE, via a \revii{pilot--tune--validate} workflow. The remainder of the section connects this objective to a scalable Dirichlet--multinomial implementation and an executable tuning recipe.

\subsection{Two-stage probability-vector randomization}\label{sec:two-stage}
We consider two-stage designs that first draw a cluster-specific treatment probability vector and then randomize units within the cluster according to that vector. Let $\Pi_j=(\Pi_j[0],\ldots,\Pi_j[M])$ denote the (random) probability vector for cluster $j$, lying on the $(M{+}1)$-simplex. The design is:
\begin{enumerate}
\item (Stage 1) Draw $\Pi_j$ independently across clusters from a researcher-chosen distribution $G$ on the simplex.
\item (Stage 2) Conditional on $\Pi_j$, assign units independently within cluster $j$:
\[
\Pr[A_{j,i}=m\mid \Pi_j]=\Pi_j[m],\qquad m\in\{0,\ldots,M\}.
\]
\revii{Equivalently, the within-cluster treatment counts satisfy
\[
(N_{j,0},\ldots,N_{j,M}) \mid \Pi_j \ \sim\ \mathrm{Multinomial}(N_j,\Pi_j).
\]}
\end{enumerate}

We analyze this Bernoulli/multinomial form of within-cluster randomization (rather than complete randomization with fixed treatment counts) because it yields tractable exposure moments (e.g., Proposition~\ref{prop:dirichlet-moments}) and matches the implementation used in our application---and, we expect, matches many relevant applications where the sample frame may not be fully enumerated prior to experiment implementation. 

\noindent\textbf{Implementation at scale.}
In large-scale deployments, sampling a distinct $\Pi_j$ for every cluster may be impractical. In our application, we approximate $G$ by a finite library $\{\pi^{(1)},\ldots,\pi^{(K)}\}$ of probability vectors constructed using low-discrepancy Sobol' sequences \citep{owen1998scrambling,sobol1967distribution}, and randomly assign each cluster $j$ to one library element; conditional on the selected $\pi^{(k)}$, unit-level assignments remain i.i.d.\ categorical as above.%
\footnote{\revii{The formulas below describe the ideal Dirichlet design. A finite implementation library only approximates that design, so prospective tuning should use the assignment moments implied by the actual library.}}

Write $p_{j[m]}(\A_j)=N_j^{-1}\sum_{i=1}^{N_j}\mathbf{1}\{A_{j,i}=m\}$ for the realized within-cluster share in arm $m$. Under this construction, $p_{j[m]}(\A_j)$ concentrates around $\Pi_j[m]$ as $N_j$ grows, and the design is entirely determined by $G$.

\begin{assumption}[Positivity]\label{ass:positivity}
For each cluster $j$ and arm $m\in\{0,\ldots,M\}$, the design satisfies $\Pr[A_{j,i}=m]>0$.
\end{assumption}
\revii{Positivity ensures that each arm can be observed, but it does not rule out an empty arm in a particular finite realization when arm counts are random. 
For the exact Dirichlet families below, marginal positivity follows from strictly positive concentration parameters and, in the asymmetric case, nonzero weights for every arm; a finite implementation must likewise give every arm positive marginal probability.}

\subsection{A single tuning parameter: Dirichlet--multinomial intermediate designs}\label{sec:dirichlet}
To obtain a one-dimensional spectrum of designs ranging from nearly cluster-uniform to nearly unit-randomized, we take $G$ to be a symmetric Dirichlet distribution:
\begin{equation}\label{eq:dirichlet}
\Pi_j \ \sim\ \mathrm{Dirichlet}(\bar\alpha,\ldots,\bar\alpha),
\qquad\text{independently for }j=1,\ldots,J.
\end{equation}
By symmetry, $\E[\Pi_j[m]]=1/(M{+}1)$ for all $m$, so marginal assignment shares are balanced across arms. Smaller $\bar\alpha$ produces more \revii{overdispersion} in $\Pi_j$ (hence more within-cluster homogeneity); larger $\bar\alpha$ concentrates $\Pi_j$ near the uniform vector (hence more within-cluster mixing). A convenient summary is the induced within-cluster correlation of the indicators $\mathbf{1}\{A_{j,i}=m\}$, which under \eqref{eq:dirichlet} takes the simple form:
\begin{equation}\label{eq:rhoT}
\rho_T(\bar\alpha)
=\mathrm{Corr}\!\left(\mathbf{1}\{A_{j,i}=m\},\mathbf{1}\{A_{j,i'}=m\}\right)
=\frac{1}{(M{+}1)\bar\alpha+1},
\qquad i\neq i',
\end{equation}
independent of $m$ by symmetry. Thus, $\bar\alpha\downarrow 0$ corresponds to $\rho_T(\bar\alpha)\uparrow 1$ (approximately cluster-uniform assignment), and $\bar\alpha\uparrow\infty$ corresponds to $\rho_T(\bar\alpha)\downarrow 0$ (approximately unit randomization with balanced mixing). This one-parameter family therefore operationalizes the continuum in Figure~\ref{fig:homogeneous_target-design}. We annotate this design as $D(\bar\alpha)$.

\noindent\textbf{\revii{Alternative marginal assignment shares.}}
\revii{Researchers may wish to use unequal or otherwise prespecified positive marginal assignment shares.} This is accommodated by replacing the symmetric Dirichlet in \eqref{eq:dirichlet} with an \emph{asymmetric} Dirichlet parameterized by a mean vector and a single concentration parameter:
\begin{equation}\label{eq:dirichlet-asym}
\Pi_j \sim \mathrm{Dirichlet}(\kappa w_0,\ldots,\kappa w_M),
\qquad w_m>0,\ \sum_{m=0}^M w_m=1,\ \kappa>0.
\end{equation}
Here $w=(w_0,\ldots,w_M)$ is the desired marginal assignment share vector while $\kappa$ controls within-cluster treatment correlation. \revii{Thus $w$ can encode any prespecified positive marginal allocation, while constraints not expressible through marginal shares would require a corresponding modification of the design.} Under \eqref{eq:dirichlet-asym}, $\E[\Pi_j[m]]=w_m$ for all $m$, and the induced within-cluster correlation of same-arm indicators remains one-dimensional:
\[
\mathrm{Corr}\!\left(\mathbf{1}\{A_{j,i}=m\},\mathbf{1}\{A_{j,i'}=m\}\right)=\frac{1}{\kappa+1},
\qquad i\neq i',
\]
independent of $m$ (and hence of $w$). The balanced design is recovered by taking $w_m\equiv 1/(M{+}1)$ and $\kappa=(M{+}1)\bar\alpha$.
All exposure-moment calculations used below extend by the same Dirichlet conjugacy argument, so we focus on the balanced case to simplify notation.

The next proposition makes the exposure environment under these designs explicit. This is the key step for turning the bias discussion into a computable function of $\bar\alpha$.

\begin{proposition}[Conditional exposure moments under Dirichlet--multinomial]\label{prop:dirichlet-moments}
Fix a cluster $j$ of size $N_j$ and the two-stage design \eqref{eq:dirichlet}. For any unit $i$ and arms $m,\ell\in\{0,\ldots,M\}$,
\begin{align}
\E\!\left[\Pi_j[\ell]\mid A_{j,i}=m\right]
&=\frac{\bar\alpha+\mathbf{1}\{\ell=m\}}{(M{+}1)\bar\alpha+1},\label{eq:condPi}\\
\E\!\left[p_{j[\ell]}(\A_j)\mid A_{j,i}=m\right]
&=\frac{1}{N_j}\mathbf{1}\{\ell=m\}
+\frac{N_j-1}{N_j}\cdot\frac{\bar\alpha+\mathbf{1}\{\ell=m\}}{(M{+}1)\bar\alpha+1}.\label{eq:condp}
\end{align}
Consequently, $\E[p_{j[\ell]}(\A_j)\mid A_{j,i}=m]\to \mathbf{1}\{\ell=m\}$ as $\bar\alpha\downarrow 0$, while $\E[p_{j[\ell]}(\A_j)\mid A_{j,i}=m]\to \mathbf{1}\{\ell=m\}/N_j + (N_j-1)/(N_j(M{+}1))$ as $\bar\alpha\uparrow\infty$.
\end{proposition}
\noindent\emph{Proof sketch.} The result follows from Dirichlet--categorical conjugacy: conditioning on a single draw $A_{j,i}=m$ updates the Dirichlet parameter for arm $m$ by $+1$, yielding \eqref{eq:condPi}; \eqref{eq:condp} then follows by linearity of expectation applied to the $N_j$ indicators. \rev{A full proof is provided in Lemma~\ref{lem:dirichlet-moments} in Appendix~\ref{app:dirichlet-moments}.}

\subsection{Bias for the GATE under intermediate designs}\label{sec:bias}
\revii{Under the equal-size and fixed-allocation conditions of Proposition~\ref{prop:dm_unbiased}, cluster-uniform randomization makes difference-in-means exactly unbiased for the GATE.}
Under intermediate designs, however, units assigned to arm $m$ are typically exposed to a non-degenerate within-cluster mix, so the conditional mean $\E[Y_{j,i}\mid A_{j,i}=m]$ generally differs from the uniform-policy mean $\E[Y_{j,i}(\bar \m_j)]$. The point of intermediate designs is that this induced bias may be small relative to the variance reduction obtained by leveraging many within-cluster units.

\revii{To make this bias channel explicit, we use the linear interference model introduced in Equation~\eqref{eq:linmod} and the model-implied GATE derived in Section~\ref{section:lim}.}

\revii{The large-sample expectation of the difference-in-means estimator under a general (non-uniform) design depends on the conditional exposure moments:}
\begin{equation}\label{eq:dm-expectation}
\E[\widehat\Psi_m^{\mathrm{DM}}]
\revii{\approx}\Big(\beta_m+\sum_{\ell=1}^M \delta_{m,\ell}\,\E[p_{j[\ell]}(\A_j)\mid A_{j,i}=m]\Big)
-\Big(\beta_0+\sum_{\ell=1}^M \delta_{0,\ell}\,\E[p_{j[\ell]}(\A_j)\mid A_{j,i}=0]\Big).
\end{equation}
\revii{The approximation replaces the expectation of a ratio with the ratio's probability limit and is asymptotically valid as the number of independent clusters grows. 
Combining Equation~\eqref{eq:dm-expectation} with Proposition~\ref{prop:dirichlet-moments}, we define $\mathrm{Bias}_m(\bar\alpha)$ for tuning as the right-hand side of \eqref{eq:dm-expectation} minus $\Psi_m^{\mathrm{GATE}}$.
This approximate bias tends to zero as $\bar\alpha\downarrow0$ (the cluster-uniform boundary), while it generally does not vanish as $\bar\alpha\uparrow\infty$ when interference is present. 
Exact unbiasedness at the cluster-uniform boundary follows from Proposition~\ref{prop:dm_unbiased} when fixed positive numbers of clusters are assigned to each arm.}

\subsection{A plug-in RMSE objective for tuning \texorpdfstring{$\bar\alpha$}{alpha-bar}}\label{sec:plugin-rmse}
To select $\bar\alpha$ in a way that is executable in practice, we approximate \eqref{eq:design-objective} with a plug-in objective computed from a pilot stage. For each candidate $\bar\alpha$ on a grid $\mathcal{A}$, define
\begin{equation}\label{eq:rmse-decomp}
\widehat{\mathrm{RMSE}}(\bar\alpha)
=\frac{1}{M}\sum_{m=1}^M\Big\{\widehat{\mathrm{Bias}}_m(\bar\alpha)^2+\widehat{\mathrm{Var}}_m(\bar\alpha)\Big\}^{1/2},
\end{equation}
where $\widehat{\mathrm{Bias}}_m(\bar\alpha)$ and $\widehat{\mathrm{Var}}_m(\bar\alpha)$ are estimator-specific. The core idea is:
\begin{itemize}
\item \textbf{Bias term.} \revii{For difference-in-means, we estimate $\{\beta_m,\delta_{m,\ell}\}$ (or a parsimonious, domain-informed subset) in the pilot and substitute them into the bias expression derived in Section~\ref{sec:bias}, using the conditional exposure moments in Proposition~\ref{prop:dirichlet-moments}. For a model-based estimator, we instead estimate its bias in design replay relative to the known model-implied GATE, using the same prespecified estimation rule that would be used in the experiment. This bias calculation is model-dependent; we therefore recommend reporting sensitivity to plausible perturbations of the fitted slopes.}
\item \textbf{Variance term.} We estimate $\widehat{\mathrm{Var}}_m(\bar\alpha)$ via Monte Carlo \emph{design replay}: simulate the full experiment under $D(\bar\alpha)$ and an outcome model fitted in the pilot, compute the estimator on each synthetic dataset, and take the empirical variance across replications; see Appendix~\ref{sec:design-replay} for details. \revii{Thus both bias and variance are matched to the estimator being tuned.}
\end{itemize}

We then choose
\begin{equation}\label{eq:alpha-star}
\bar\alpha^\star\ \in\ \arg\min_{\bar\alpha\in\mathcal{A}}\ \widehat{\mathrm{RMSE}}(\bar\alpha),
\end{equation}
and run the main experiment using $D(\bar\alpha^\star)$. The same procedure can be applied to compare \emph{(design, estimator)} pairs by computing \eqref{eq:rmse-decomp} for each estimator of interest.

\subsection{Pilot--tune--validate workflow}\label{sec:workflow}
We summarize the design-selection procedure as a pilot--tune--main-run--validate workflow intended for large-scale experiments. The pilot is exploratory: its purpose is to learn the ingredients that govern \eqref{eq:rmse-decomp}. 
The tuned main run then uses the chosen $\bar\alpha^\star$ for high-precision arm ranking or selection. Finally, when unbiased estimation and valid coverage for the uniform-policy effect are required, a cluster-uniform validation experiment can confirm the selected arm(s).

\begin{enumerate}
\item \textbf{Pilot (explore).} \revii{Run one concurrent randomized pilot, partitioning a set of independent clusters across several prespecified $\bar\alpha$ values.
Independence across clusters follows from our i.i.d. superpopulation sampling assumption, while partial interference rules out treatment spillovers across clusters.
Provided the pilot design generates sufficient variation in realized within-cluster treatment shares, estimate} (i) the cluster size distribution $\{N_j\}$, \revii{(ii) either the variance components $(\sigma^2,\tau^2)$ or the total marginal variance $\sigma^2+\tau^2$ together with the intra-cluster outcome correlation $\rho_u$,} and (iii) a parsimonious interference summary (e.g., selected slopes $\delta_{m,\ell}$). \revii{The full linear model contains $(M+1)(M+1)$ mean parameters: $M+1$ direct-treatment intercepts $\beta_m$ and $M(M+1)$ exposure slopes $\delta_{m,\ell}$. A pilot need not estimate all of them; domain restrictions, pooling, or regularization may be essential in a many-arm experiment. Even if all were known, Equation~\eqref{eq:linmod} would give the GATE only under correct model specification, and estimated parameters retain sampling uncertainty.}
\revii{Assess the stability of the estimated RMSE curve under re-estimation (e.g., bootstrap over clusters) and sensitivity analysis. A boundary minimizer is not itself a failure: expand the candidate grid only when the boundary result appears to reflect grid truncation; otherwise retain a stable endpoint solution. Appendix~\ref{app:workflow-guidance} gives further guidance.}
\item \textbf{Tune (minimize predicted RMSE).} For each $\bar\alpha\in\mathcal{A}$, compute $\widehat{\mathrm{RMSE}}(\bar\alpha)$ using \revii{(i) closed-form conditional exposure moments from Proposition~\ref{prop:dirichlet-moments} for the difference-in-means bias term and (ii) estimator-specific bias and variance from Monte Carlo design replay under $D(\bar\alpha)$ where required. The replay generates new assignments and synthetic outcomes from the pilot-fitted working model.}
Select $\bar\alpha^\star\in\arg\min_{\bar\alpha\in\mathcal{A}}\widehat{\mathrm{RMSE}}(\bar\alpha)$ and examine sensitivity to slope and variance-component uncertainty.
\item \textbf{Main run (rank/select).} Run the main multi-arm experiment using $D(\bar\alpha^\star)$ \revii{on fresh clusters} and select promising arm(s) using the \revii{prespecified} decision rule (e.g., best estimated performance on the primary metric, potentially under constraints).
\item \textbf{Validate (optional).} If unbiased GATE estimation and correct coverage are required for the final claim, validate the selected arm(s) in a cluster-uniform experiment and report the design-based GATE estimate and uncertainty.
With low RMSE estimates of the effects already obtained, power analysis can determine the necessary sample size for this validation.
\end{enumerate}
}

\section{Application: Facebook video playback experiments}\label{section:fb}
\revii{In our study, units are user--video impressions grouped into user sessions. We compare session-level randomization, under which every video impression in a user session receives the same arm, with within-session randomization, under which video impressions within a user session may receive different arms.}

\revii{Apart from the level of randomization, nothing else was changed about the delivery of content to users' feeds in the course of these experiments.}
This means that the same ranking algorithms delivered content to users' feeds across each treatment group.
The algorithm at the time of this experiment did no online adaptation in response to users' consumption of videos.
The content of videos is largely irrelevant to the speed at which videos would load and, therefore, to their ability to be viewed by users.
That said, behavioral changes as a result of differing performance of the video player \revii{are} nearly certain.
For example, if \revii{a video played very poorly}, then users would likely spend less time viewing it (which could result in concomitant greater or lesser overall time spent on the platform depending on their preferences around content).
These behavioral changes are not the subject of our experiment and we focus attention on a direct measure of the performance of the video players by examining the number of stalls experienced in the course of viewing.

\begin{figure}[!ht]
\centering
\includegraphics[width=.9\textwidth]{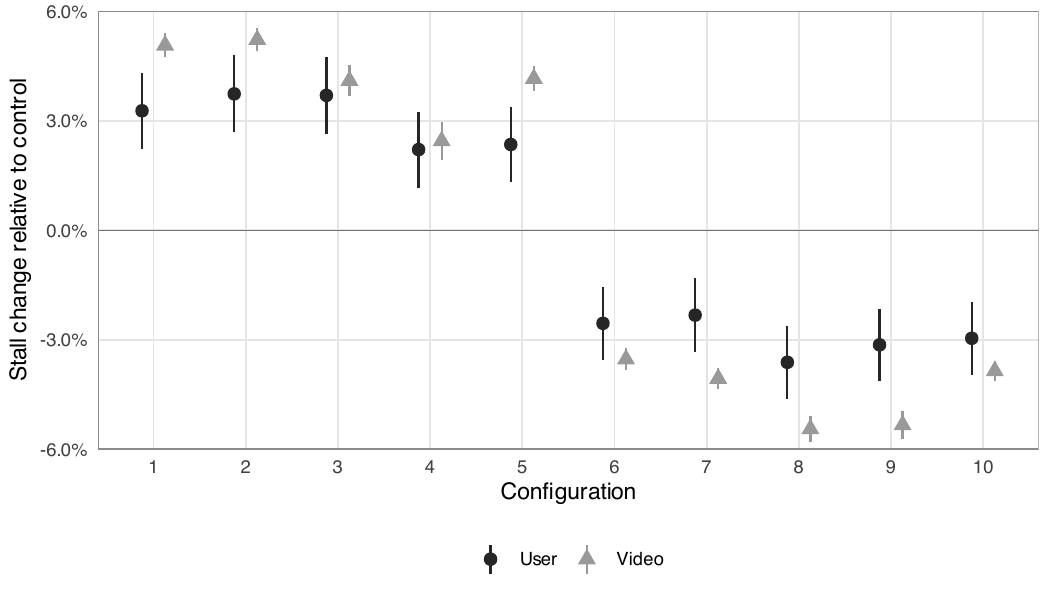}
\caption{\textbf{\revii{Stall-count effect estimates under user--level and user--video randomization.}} \revii{Dark circles show user--level randomization, and gray triangles show user--video randomization. Points are percentage changes relative to the design-specific control; bars extend 1.96 reported delta-method standard errors on either side of each estimate. Configurations 1--5 are low-data variants and configurations 6--10 are high-data variants.}}
\label{fig:homogeneous_cupr-ulr}
\end{figure}

\revii{Figure~\ref{fig:homogeneous_cupr-ulr} illustrates estimates from user--level randomization and user--video randomization experiments for one representative outcome variable, stall counts.}
Stalls are a negative user experience, so the goal is to reduce them.
\revii{Dark circles denote user--level randomization, under which all videos for a user received a common arm, and gray triangles denote user--video randomization, under which assignment was keyed by the user--video pair.}
\revii{Both experiments were run over four days.}
\revii{Ten treatment conditions, numbered 1--10 in Figure~\ref{fig:homogeneous_cupr-ulr}, were assigned in addition to the control, the status quo configuration of engineering parameters. Configurations 1--5 are low-data variants and configurations 6--10 are high-data variants, relative to the control condition.}
\rev{The ten non-control variants were organized as five matched \revii{pairs of low- and high-data configurations} that primarily vary buffering intensity (and hence data usage), which can reduce stalls but also increase within-session competition for bandwidth. Across pairs, remaining buffer-related parameters were varied to span multiple operating points within each data-usage level; see Appendix~\ref{app:fb-variants} for details.}

\revii{The user--level experiment supplies a uniform-policy benchmark because it assigns a common arm to all videos and sessions for a user. 
Its estimates are much less precise than those from the user--video randomization experiment: for stall counts, the average arm-specific delta-method standard error is approximately three times larger under the user--level design. }
\revii{The user--video effects are measured to be around one and a half times more extreme than their user--level counterparts. Thus, the user--video randomization experiment had substantially smaller reported standard errors in this comparison, while the systematic cross-design differences are consistent with interference-induced bias relative to the user--level uniform-policy benchmark.
(Note that estimates are percentage changes relative to each experiment's own control, and so do not directly reflect the additive GATE in Equation~\eqref{eq:gate}; however, these contrasts provide descriptive information about the relative performance of the two designs.)}

\rev{No analyst would want to use very precise effect estimates that are highly biased; however, high variance is also problematic in practice because it limits optimization and arm ranking.
In practice, the \revii{video player} is often subject to many simultaneous experiments, as ensuring that it is well-tuned is critical to the smooth operation of the platform.
Thus, reducing variance and increasing throughput of experimentation is an important consideration.}
\revii{This motivates intermediate two-stage designs that trade off these channels and a workflow that uses lower-level designs for screening and tuning followed, when unbiased estimation is required, by user--level validation. }
%Across the 13 outcomes measured, effects were only estimated as different from zero at standard significance levels for about 30\% of the estimates. Estimates from the unit-randomized experiment were much more precise, and also more extreme. Treatment effects were estimated as different from zero at standard significance levels across over 95\% of the estimates.

The differences between the two designs are also systematic by treatment type and are consistent with the proposed direction of design-induced bias: \revii{high-data treatments} were estimated as being even more effective under user--video randomization than under user--level randomization, while \revii{low-data treatments} were estimated as being even less effective. 
This may be based on the nature of data usage: when all videos for a user---and hence all videos within each session---are assigned a \revii{low-data configuration}, there is relatively little competition for bandwidth.
When a given video receives a \revii{low-data configuration} under \revii{user--video randomization} and the other videos in its session receive a mixture of \revii{low- and high-data configurations}, increased competition for bandwidth may make the low-data video perform worse than under user--level uniform assignment. 
Conversely, when all videos for a user are assigned a \revii{high-data configuration}, there is a high degree of competition for bandwidth. 
Under \revii{user--video randomization}, a high-data video may face less competition when the other videos in its session receive a mixture of low- and high-data configurations, making it appear more effective than under user--level uniform assignment. 
\rev{That is, for some arms the induced bias may be modest relative to the precision gains from lower-level randomization.}
This creates a finite-sample trade-off: intermediate or unit-level designs can have acceptable, and sometimes lower, effect-estimation RMSE when the bias is small, and this greater precision may facilitate ranking configurations. 
\revii{In our motivating setting, the operational goal is to rank and optimize configurations using experimentation, so precise effect estimates are useful.}

\rev{We used data from the \revii{user--level and user--video randomized experiments} to estimate intra-cluster correlations of errors and the mean and variance of the number of videos per session, as clusters were not of uniform size.}
Based on domain knowledge, we estimated interference as a linear effect of \revii{the proportion of high-data videos in a session}. 
Table~\ref{table:homogeneous_fbests} reports estimates of parameters for the stall count outcome, averaged over all treatments. 
\revii{We used these quantities to inform simulations and select a value of $\bar \alpha$---equivalently, $\rho_T(\bar\alpha)$---designed to reduce RMSE, and conducted a later experiment following an adapted version of the Dirichlet--multinomial procedure described above.}
The video playback configurations we sought to optimize had five continuous parameters; in the \revii{second-stage experiment}, we tested 30 unique configurations, with an approximately balanced distribution of \revii{high- and low-data configurations}.

\begin{table}[!ht]
\centering
\begin{tabular}{lllllll}
\hline
 & $\rho_u$ & SEs, user--level & SEs, user--video & Magnitude ratio & $\Var(N_j)$ & $\bar{N}$ \\
 \hline
 & 0.1  & 0.00522  & 0.00180 &  1.46 & 282.31 & 9.25\\
\hline
\end{tabular}
\caption{\textbf{We estimate parameters that inform experimental design from the experiments described in Figure~\ref{fig:homogeneous_cupr-ulr}.}}
\label{table:homogeneous_fbests}
\begin{flushleft}
\singlespacing
\small
\revii{Columns report the estimated intra-cluster correlation of errors, $\rho_u$; average standardized SEs under the user--level and user--video randomized experiments; the ratio of the average absolute user--video effect magnitude to the corresponding user--level magnitude; and the variance $\Var(N_j)$ and mean $\bar N=J^{-1}\sum_{j=1}^J N_j$ of cluster size.}
\end{flushleft}
\end{table}
\rev{In the large-scale online setting it was not practical to assign a unique draw from a Dirichlet distribution to every session in the sample, so we implemented randomization by \revii{precomputing} a low number of quasi-random draws from the Dirichlet distribution which were then deployed at scale.}
These draws were selected based on low-discrepancy Sobol' sequences \citep{sobol1967distribution} to more evenly cover the sample space as compared to alternative \revii{pseudorandom} sampling procedures.
Within clusters, treatment was then randomly assigned according to the relevant multinomial distribution.

\rev{We used this approach to optimize the engineering configurations for video playback, which we validated in a uniform session-level experiment.}
By reducing the variance of effect estimates through design selection, we were then able to use constrained Bayesian optimization to select an optimal configuration from the continuous space of engineering parameters.
We direct interested readers to \citet{letham2019noisyei} for more on this optimization process. %Optimization of noisy outcome variables in such a complicated space of potential parameters was infeasible in the absence of this sort of variance-reducing design.
The proposed configuration resulted in a reduction in stall counts without otherwise degrading measured outcomes. % eventually get more specific

\section{Simulations}\label{section:sim}
To demonstrate a range of scenarios, we develop simulations with varying intra-cluster correlation of errors and magnitude of interference.
We simulate an experiment with 100 clusters, each with $N = 50$.
Errors are normally distributed with mean zero and $\sigma^2$ fixed at one; $\tau^2$ varies following $\rho_u/(1-\rho_u)$ according to the values in Table~\ref{table:homogeneous_param}.%
\revii{(Under this parameterization, increasing $\rho_u$ also increases total error variance.)}
Magnitude of interference is defined by $c$, values of which are also listed in the Table. \revii{The interference coefficients in the data-generating process are $c\delta_{m,\ell}$.}
\revii{We run 1,000 simulations at each combination of $\rho_u \times c \times \kappa$ parameter values, where $\kappa=(M{+}1)\bar\alpha$.}

We consider a setting with three treatment conditions, motivated by the Facebook application discussed above, where there is a control condition, as well as a ``high'' treatment condition (treatment condition one) and a ``low'' treatment condition (treatment condition two).
Compared to the \rev{uniform} control setting, \rev{uniform} treatment under the ``high'' treatment condition results in the highest outcomes; \rev{uniform} treatment under the ``low'' treatment condition results in lowest outcomes.
\rev{Fixing direct treatment, outcomes are linearly increasing with the proportion of ``high'' treatment assignments in a cluster, and linearly decreasing with proportion of ``low'' treatment assignments in a cluster.}
We let the \revii{data-generating} process follow the linear model with slope effects in Equation~\ref{eq:linmod}, specialized to $M=2$ (control plus two treatments). We vary the interference magnitude scalar $c$, where $c=0$ corresponds to no interference and larger $c$ increases the strength of interference linearly. Consequently, expected potential outcomes under \rev{uniform} treatment are 5 under control, $7.5 + c $ under the ``high'' treatment, and $2.5-2.5c$ under the ``low'' treatment.

\begin{table}[H]
\centering
{\small
\setlength{\tabcolsep}{4pt}
\begin{tabular}{@{}lp{0.78\textwidth}@{}}
\toprule
\textbf{Parameter(s)} & \textbf{Value(s)} \\
\midrule
$\beta$ & $(\beta_0,\beta_1,\beta_2)=(5,\,7.5,\,2.5)$ \\
$\delta$ & $\big(\delta_{m,\ell}\big)_{m=0,1,2;\,\ell=1,2}
=\left(\begin{smallmatrix}
\rev{0.5} & \rev{-0.5}\\
\rev{1} & \rev{-1}\\
\rev{2.5} & \rev{-2.5}
\end{smallmatrix}\right)$ \\
$\sigma^2$ & $1$ \\
$\rho_u$ & $\{0,\,0.1,\,0.3,\,0.5,\,0.8\}$ \\
$c$ & $\{0,\,0.1,\,0.5,\,1\}$ \\
$\kappa=(M{+}1)\bar{\alpha}$ & $\{0.001,\,0.01,\,0.05,\,0.1,\,0.2,\,\ldots,\,0.9,\,1,\,2,\,3,\,10,\,1000\}$ \\
\bottomrule
\end{tabular}
}
\caption{\textbf{Simulation DGP coefficients (Equation~\ref{eq:linmod} with $M=2$) and parameter grids.}}
	\label{table:homogeneous_param}
\end{table}

We estimate the GATE both under the correctly specified linear model and the \revii{difference-in-means estimator}, using OLS with \revii{HC1 cluster-robust variance estimates. When a target contrast remains estimable in a rank-deficient fit, the simulation code computes that contrast and its standard error directly; for full-rank fits, the variance estimate is equivalent to the implementation in the \texttt{sandwich} package \citep{zeileis2006object-oriented, berger2017various} in \texttt{R}. Both uniform-policy contrasts were estimable in every production replication, even when individual exposure coefficients were not separately identified.}

\revii{For varying levels of intra-cluster correlation of errors, $\rho_u$, and magnitude of interference, $c$, Appendix Table~\ref{table:homogeneous_sims_summary} reports, for each estimator, the value of $\rho_T$ on the simulated grid with the lowest obtained RMSE. 
Selected results at three representative values of $\rho_T(\bar \alpha)$ are reported in Appendix Table~\ref{table:homogeneous_sims}.}

\begin{figure}[H]
\centering
  \includegraphics[width=0.95\textwidth]{\string 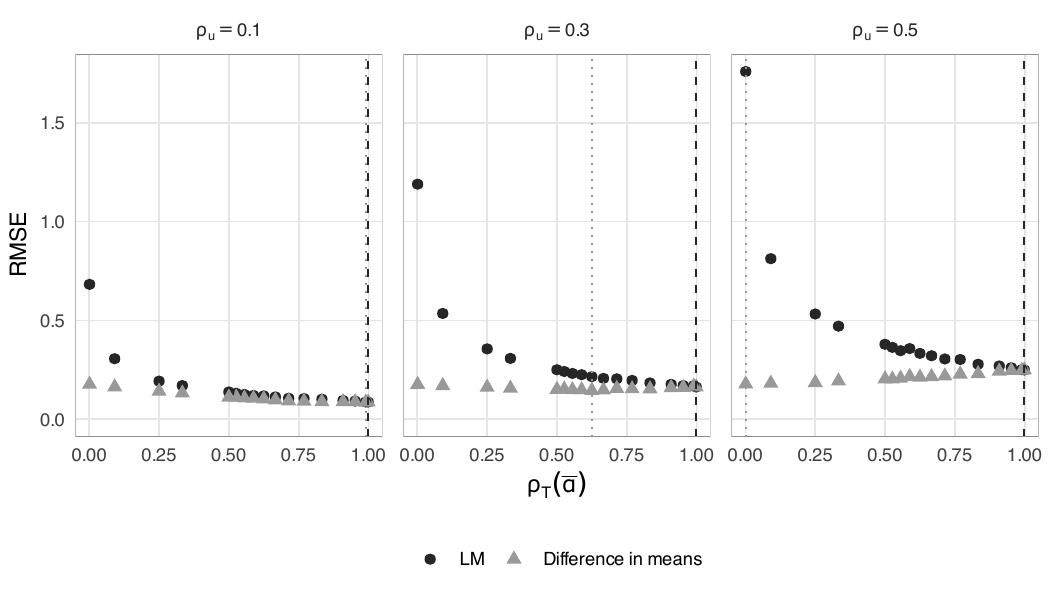}
\caption{\textbf{For the linear model, RMSE generally decreases with $\rho_T(\bar\alpha)$; for the difference-in-means estimator, the minimum simulated RMSE may occur at an interior value of $\rho_T(\bar\alpha)$.}
\revii{Dark circles plot linear-model RMSE and gray triangles plot difference-in-means RMSE against the treatment ICC $\rho_T(\bar\alpha)$.}
Magnitude of interference, $c$, is fixed at 0.1; intra-cluster correlation of errors $\rho_u$ varies over $\{0.1, 0.3, 0.5\}$.
\revii{Vertical lines mark each method's minimum-RMSE value on the simulated grid.}}
\label{fig:homog_sim03}
\end{figure}

\revii{In this setting, the GATE is generally estimated with higher RMSE under OLS with the correctly specified linear model as compared to the difference-in-means estimator when interference is low (here, $c \le .1$); the differences are generally modest by $c=.5$.}
This is largely attributable to the greater variance of the linear estimator.
However, the linear estimator generally has lower bias and \revii{approximately nominal coverage}, while the difference-in-means estimator \revii{can have} poor to very poor coverage in the presence of interference when randomization is not at the cluster level, as demonstrated in Appendix Table~\ref{table:homogeneous_sims}.
\revii{At $c=1$, both estimators favor near cluster-uniform assignment and perform similarly.}
\revii{These simulations were all conducted with fixed $N$ and $J$. For the difference-in-means estimator under a fixed design that is not cluster-uniform and has nonzero asymptotic bias, that bias will eventually dominate the decreasing variance as the number of independent clusters grows.}

For the linear estimator, in this setting to minimize RMSE for the GATE it is nearly always preferable to set the design as close to cluster-level randomization as possible ($\rho_T(\bar \alpha) \approx 1$).
%http://madrury.github.io/jekyll/update/statistics/2016/07/20/lm-in-R.html

For the difference-in-means estimator, when there are high levels of interference (here, $c\ge.5$), the minimum RMSE is obtained with designs that randomize closer to the cluster level.
When there is no interference but errors are correlated, it is preferable to set designs with randomization closer to the unit level.
\revii{When there is no interference and no intra-cluster correlation of errors, the choice of $\bar \alpha$ has little practical effect in this simulation, although small finite-sample differences can remain because random arm counts affect the variance of the sample means.} When there is an intermediate level of interference (here, $c = .1$), the design trade-offs become evident at different levels of $\rho_u>0$, as is illustrated in Figure~\ref{fig:homog_sim03}.
\revii{For $\rho_u \in \{0.1, 0.3, 0.5 \}$, the lowest obtained difference-in-means RMSE occurred at $\rho_T(\bar\alpha)=0.990$, $0.625$, and $0.001$, respectively. Thus the selected design moves from near cluster-uniform assignment, through an interior design, to near unit randomization as $\rho_u$ increases.}

\section{Conclusion}
Researchers may wish to estimate the average effect of a \rev{uniform policy} change in the presence of interference, and so will be concerned with both experimental design and estimator.
When using RMSE as a metric with a finite sample, it may be preferable to use the difference-in-means estimator, even when the linear model is correctly specified.
\revii{In many applied settings, difference-in-means estimators may also be more practical to deploy at scale.}

As a general guideline when using the difference-in-means estimator, when interference is low or absent, designs closer to unit randomization (lower $\rho_T$) tend to have lower RMSE by reducing variance, with the gains largest when outcomes are correlated within clusters ($\rho_u>0$). When interference is substantial, designs closer to cluster-uniform randomization (higher $\rho_T$) tend to have lower RMSE by avoiding bias. In intermediate regimes, interior values of $\rho_T$ can be RMSE-optimal.

%Setting the design to account for the level of interference and intra-cluster correlation of errors can further decrease RMSE.

\revii{However, coverage for the GATE generally will not be correct under intermediate assignment when interference is present. If correct coverage is necessary for the final claim, the selected arm should be validated with cluster-level randomization.} If the objective of analysis is to test for or measure interference, an approach more similar to that proposed by \cite{baird2018optimal}, Propositions 1 and 2, would be recommended, using a model informed by domain knowledge and minimizing standard errors on estimates of slope effects.

%% file: appendix.tex
% \section{}\label{homog_appendix:proofs}

% \subsection{Unbiasedness of the OLS estimate}\label{homog_appendix:unbiased}
% [TKTK]

% \subsection{Intra-cluster correlation of treatment as a function of $\bar\alpha$}\label{homog_appendix:icct}

% [TKTK]

% \newpage

\appendix

\rev{
\section{Worked calculations for the toy example}\label{app:toy}

This appendix records the bias and variance calculations for the simple difference-in-means estimator
$\widehat{\Delta}_m=\overline{Y}_{A=m}-\overline{Y}_{A=0}$ in the toy example from the Introduction.
\revii{The estimand for arm $m\in\{1,2\}$ is the uniform-policy contrast $\Psi_m^{\mathrm{GATE}}=\E[Y_{j,i}(\bar\m_j)-Y_{j,i}(\bar{\bm{0}}_j)]$.}

\revii{\noindent\textbf{Toy setup.}
There are $J=3$ user sessions (clusters), each containing $N=6$ videos (units). Arm 0 is control, arm 1 is high-data, and arm 2 is low-data. Let $n_{j\ell}=\sum_{i=1}^N \mathbf{1}\{A_{j,i}=\ell\}$ be the number of videos assigned to arm $\ell$ in session $j$. Each session has its own fixed baseline mean, $\mu_j$. The expected outcomes are
\begin{align*}
\E[Y_{j,i}\mid A_{j,i}=0,\A_j]&=\mu_j,\\
\E[Y_{j,i}\mid A_{j,i}=1,\A_j]&=\mu_j+0.5-0.05\,n_{j2},\\
\E[Y_{j,i}\mid A_{j,i}=2,\A_j]&=\mu_j+0.75+0.05\,n_{j1}.
\end{align*}
For this toy calculation, outcomes are taken to equal these conditional means. Under uniform rollout, the same average session baseline appears under every arm and therefore cancels from the treatment--control contrasts. The two GATEs are consequently $\Psi_1^{\mathrm{GATE}}=0.5$ and $\Psi_2^{\mathrm{GATE}}=0.75$. For each design below, its three treatment-count patterns are assigned at random to the three sessions. We use $\E_D$ and $\operatorname{Var}_D$ for the mean and variance over this random assignment, keeping the three session baselines fixed.}

\subsection{Bias under interference (mixed exposure vs.\ GATE)}
\revii{\noindent\textbf{Cluster-uniform assignment.}
The three session patterns are $(6,0,0)$, $(0,6,0)$, and $(0,0,6)$. There is no mixing within a session, so the interference terms are zero. The treatment and control groups generally come from sessions with different baseline means, but each session is equally likely to receive each arm. The baseline difference is therefore zero on average, giving
\[
\E_D[\widehat\Delta_1]=0.5,
\qquad
\E_D[\widehat\Delta_2]=0.75.
\]
Thus cluster-uniform assignment is unbiased for both GATEs.

\noindent\textbf{Balanced complete randomization.}
Every session has $(n_{j0},n_{j1},n_{j2})=(2,2,2)$. Because all three arms use the same number of videos from every session, the session baselines cancel exactly. Every high-data video is exposed to two low-data videos, and every low-data video is exposed to two high-data videos. Therefore
\[
\widehat\Delta_1=0.5-0.05(2)=0.4,
\qquad
\widehat\Delta_2=0.75+0.05(2)=0.85.
\]

\noindent\textbf{Intermediate design.}
The three patterns are $(4,1,1)$, $(1,4,1)$, and $(1,1,4)$. For either treatment arm, five of its six videos share a session with one video from the other treatment arm, while the remaining video shares a session with four videos from the other treatment arm. The session baselines do not cancel for every particular assignment of the patterns, but their contribution is zero on average because every session is equally likely to receive each pattern. Hence
\begin{align*}
\E_D[\widehat\Delta_1]
&=\{5(0.5-0.05\cdot1)+(0.5-0.05\cdot4)\}/6=0.425,\\
\E_D[\widehat\Delta_2]
&=\{5(0.75+0.05\cdot1)+(0.75+0.05\cdot4)\}/6=0.825.
\end{align*}
Both mixed designs are biased for the GATEs, but the intermediate design has less bias than balanced complete randomization.}

\subsection{\revii{Design-based variance from fixed session baselines}}
\revii{Let
\[
\bar\mu=\frac{1}{J}\sum_{j=1}^J\mu_j,
\qquad
S_\mu^2=\frac{1}{J-1}\sum_{j=1}^J(\mu_j-\bar\mu)^2
\]
be the average and variance of the three fixed session baselines.

Under \textbf{cluster-uniform assignment}, each treatment--control estimate compares the baseline of one randomly selected session with the baseline of another. The variance of this baseline difference is $2S_\mu^2$.

Under \textbf{balanced complete randomization}, every arm receives the same number of videos from every session. The baseline contribution cancels exactly, so its variance is zero.

Under the \textbf{intermediate design}, the treatment--control contrast places half as much weight on the difference between two session baselines as cluster-uniform assignment does. Scaling a difference by one half scales its variance by one quarter, so the contribution is
\[
\frac14(2S_\mu^2)=\frac12S_\mu^2.
\]}
}

\rev{
\section{Monte Carlo design replay for the variance term}\label{sec:design-replay}
To obtain $\widehat{\mathrm{Var}}_m(\bar\alpha)$ for candidate designs without additional algebra, we use a Monte Carlo \emph{design replay} procedure that simulates the full experiment under the candidate design and an outcome model fitted in the pilot. \revii{This is a prospective simulation, not a claim that observed pilot outcomes can be held fixed and reanalyzed under counterfactual assignments. No cluster is experimentally rerun.} The replay is used only to tune $\bar\alpha$ (i.e., to compare predicted RMSE across designs).

\noindent\textbf{Inputs.}
Fix: (i) planned cluster count $J$ and cluster sizes $\{N_j\}_{j=1}^J$ (or an empirical distribution of $N_j$ from the pilot), (ii) a candidate design parameter $\bar\alpha$ (or $\kappa$ in the unbalanced version \eqref{eq:dirichlet-asym}), (iii) the pilot-fitted linear interference model parameters $(\widehat\beta,\widehat\delta)$ from \eqref{eq:linmod}, and \revii{(iv) either pilot estimates of the variance components $(\widehat\sigma^2,\widehat\tau^2)$ or the estimated total marginal variance $\widehat v=\widehat\sigma^2+\widehat\tau^2$ together with $\widehat\rho_u$ from Assumption~\ref{ass:homoskedasticity}. In the latter parameterization, $\widehat\tau^2=\widehat\rho_u\widehat v$ and $\widehat\sigma^2=(1-\widehat\rho_u)\widehat v$.} Let $B$ denote the number of Monte Carlo replications.

\noindent\textbf{Step 1: simulate assignment under $D(\bar\alpha)$.}
For each replication $b=1,\ldots,B$ and each cluster $j$:
\begin{enumerate}
\item Draw a cluster probability vector $\Pi_j^{(b)}\sim \mathrm{Dirichlet}(\bar\alpha,\ldots,\bar\alpha)$ (or the asymmetric Dirichlet \eqref{eq:dirichlet-asym} when using unbalanced marginal shares). \revii{When evaluating a finite implementation library, instead sample $\Pi_j^{(b)}$ from the actual library using its deployment probabilities.}
\item Draw treatment counts
\[
(N_{j,0}^{(b)},\ldots,N_{j,M}^{(b)})\mid \Pi_j^{(b)} \sim \mathrm{Multinomial}(N_j,\Pi_j^{(b)}).
\]
\item Assign exactly $N_{j,m}^{(b)}$ units in cluster $j$ to arm $m$ by complete randomization within cluster.
\end{enumerate}
Compute realized within-cluster shares $p_{j[\ell]}^{(b)}=N_{j,\ell}^{(b)}/N_j$ for all $\ell$. \revii{Record whether every arm entering the design objective is nonempty. Retain valid assignments for the conditional RMSE calculation and redraw the full experiment-wide assignment otherwise. The fraction of initially valid draws estimates the probability that a single assignment draw has positive counts in every required arm and should be reported for each candidate design; candidates with a non-negligible probability of an invalid initial assignment should not be selected.}

\noindent\textbf{Step 2: simulate outcomes under the pilot-fitted model.}
Generate outcomes according to the linear interference model with clustered errors:
\begin{equation}\label{eq:design-replay-DGP}
Y^{(b)}_{j,i}
=\sum_{m=0}^M \widehat\beta_m\,\mathbf{1}\{A^{(b)}_{j,i}=m\}
+\sum_{m=0}^M\sum_{\ell=1}^M \widehat\delta_{m,\ell}\,p_{j[\ell]}^{(b)}\mathbf{1}\{A^{(b)}_{j,i}=m\}
+u^{(b)}_j+e^{(b)}_{j,i},
\end{equation}
where $u^{(b)}_j\sim \mathcal{N}(0,\widehat\tau^2)$ and $e^{(b)}_{j,i}\sim \mathcal{N}(0,\widehat\sigma^2)$ are mutually independent across $j$ and $i$.

\noindent\textbf{Step 3: compute the estimator.}
Using the synthetic data $\{A^{(b)}_{j,i},Y^{(b)}_{j,i}\}$, compute the estimator of interest for each arm $m=1,\ldots,M$. For example:
\begin{itemize}
\item DM: compute $\widehat\Psi_m^{\mathrm{DM},(b)}$ using \eqref{eq:DM} on $\{A^{(b)},Y^{(b)}\}$.
\item \revii{LM}: refit \eqref{eq:linmod} by OLS on $\{A^{(b)},Y^{(b)}\}$ and compute the plug-in $\widehat\Psi_m^{\mathrm{LM},(b)}=(\widehat\beta^{(b)}_m+\widehat\delta^{(b)}_{m,m})-\widehat\beta^{(b)}_0$ \revii{whenever this linear combination is estimable. If a replay draw does not identify the target contrast, a prespecified alternative estimator must be used; our implementation uses the difference-in-means contrast in Equation~\eqref{eq:DM}. This safeguard was never invoked in the reported simulations.}
\end{itemize}

\noindent\textbf{Step 4: estimate variance (and RMSE).}
For each arm $m$, estimate the variance under $D(\bar\alpha)$ by the Monte Carlo variance:
\[
\widehat{\mathrm{Var}}_m(\bar\alpha)
=\frac{1}{B-1}\sum_{b=1}^B\left(\widehat\Psi_m^{(b)}-\frac{1}{B}\sum_{b'=1}^B \widehat\Psi_m^{(b')}\right)^2.
\]
\revii{For difference-in-means, combine this variance with the plug-in bias from Proposition~\ref{prop:dirichlet-moments} and \eqref{eq:dm-expectation}. Because those exposure moments are derived before applying the empty-arm redraw rule, they approximate the bias across the retained assignments; the discrepancy vanishes as the probability of an initially invalid assignment approaches zero. For a model-based estimator, estimate bias directly across the retained replay replications relative to the known model-implied GATE, using the same prespecified estimation rule on every replay draw. We then combine the estimator-specific bias and variance to obtain $\widehat{\mathrm{RMSE}}(\bar\alpha)$ in \eqref{eq:rmse-decomp} and select $\bar\alpha^\star$ in \eqref{eq:alpha-star}.}

\noindent\textbf{Variance reduction for tuning.}
\revii{To stabilize comparisons across $\bar\alpha$ values, we suggest using \emph{common random numbers}: holding the Monte Carlo draws $\{u^{(b)}_j,e^{(b)}_{j,i}\}$ and the random seeds fixed across candidate $\bar\alpha$ values can reduce, though not eliminate, Monte Carlo noise in pairwise design comparisons.}

\noindent\textbf{Unequal cluster sizes.}
For clarity, much of the exposition begins under the equal-size simplification $N_j\equiv N$. In applications cluster sizes may vary (\revii{e.g., sessions contain different numbers of videos}). All conditional-moment expressions used above remain valid conditional on $N_j$ after replacing $N$ by $N_j$. \revii{Because the GATE is unit-weighted, we combine any cluster-size-specific bias moment $g(N_j)$ using the size-weighted version}
\[
\widehat{\E}_{\text{unit}}[g(N)] \;=\; \frac{\sum_{j=1}^J N_j\, g(N_j)}{\sum_{j=1}^J N_j}.
\]
\revii{We apply this weighting when computing $\widehat{\mathrm{Bias}}_m(\bar\alpha)$. The Monte Carlo design replay uses the complete planned or empirical size vector $\{N_j\}_{j=1}^J$ directly when estimating $\widehat{\mathrm{Var}}_m(\bar\alpha)$.}

\subsection{Pilot sizing and practical guidance}\label{app:workflow-guidance}

The pilot is not intended to deliver a final unbiased GATE estimate; it is intended to identify whether the RMSE-minimizing design is near the \revii{cluster-uniform} extreme, the unit-randomized extreme, or in between.
\revii{Rather than imposing a universal stopping rule, we recommend assessing whether the estimated RMSE curve and its minimizer are stable across repeated re-estimation (e.g., bootstrap resampling of clusters) and plausible perturbations of the working model. A boundary minimum may be the correct endpoint solution; expand the candidate grid only when the fitted curve suggests that the current grid truncates the relevant range.}
If the minimum occurs near $\rho_T\approx 1$, the workflow collapses to a standard \revii{cluster-uniform} experiment; if it occurs near $\rho_T\approx 0$ and the estimated bias term is negligible, the workflow collapses to unit randomization.

Intermediate designs are most useful when (i) clusters are moderately large (so there is substantial within-cluster information to exploit), (ii) outcomes exhibit nontrivial within-cluster correlation, and (iii) interference exists but is not so strong that even modest deviations from \revii{cluster-uniform assignment} create unacceptable bias.
In these regimes, small increases in within-cluster mixing can yield large variance reductions, and the RMSE criterion \eqref{eq:rmse-decomp} can favor $\bar\alpha^\star$ strictly between the cluster- and unit-randomized extremes.
 
If the analysis goal is unbiased estimation with valid confidence intervals for the uniform-policy effect (rather than optimization or ranking), then \revii{cluster-uniform} assignment remains the default recommendation.
In that case, intermediate designs can still be used as a screening tool in an earlier optimization phase, followed by a confirmatory \revii{cluster-uniform} experiment for final inference.
A simple decision rule is to proceed with intermediate designs only when the predicted bias component is small relative to the predicted standard deviation for a range of $\bar\alpha$ values that meaningfully reduce variance; otherwise, use intermediate designs only for screening and rely on \revii{cluster-uniform} validation for the final claim.
}

\rev{
\section{Proofs}\label{app:proofs}

\subsection{Unbiasedness of difference-in-means for the GATE}\label{app:proof-dm-unbiased}
\revii{\emph{Proof.} Under cluster-uniform assignment and equal cluster size $N$, define the cluster-average potential outcomes $\overline Y_j(\bar\m_j)=N^{-1}\sum_iY_{j,i}(\bar\m_j)$ and $\overline Y_j(\bar{\bm{0}}_j)=N^{-1}\sum_iY_{j,i}(\bar{\bm{0}}_j)$. Complete randomization assigns a fixed positive number of clusters to each arm, so consistency and randomization imply
\begin{align}
\E_D\!\left[\widehat\Psi_m^{\mathrm{DM}}\mid\mathcal S\right]
&=\frac{1}{J}\sum_{j=1}^J\left\{\overline Y_j(\bar \m_j)-\overline Y_j(\bar{\bm{0}}_j)\right\}.
\end{align}
Taking the superpopulation expectation and using partial interference gives
\begin{align}
\E_{\mathcal S}\!\left[\frac{1}{J}\sum_{j=1}^J\left\{\overline Y_j(\bar \m_j)-\overline Y_j(\bar{\bm{0}}_j)\right\}\right]
&=\E\left[Y_{j,i}(\bar \m_j)-Y_{j,i}(\bar{\bm{0}}_j)\right]
=\Psi_m^{\mathrm{GATE}}.
\end{align}}
}

\rev{
\subsection{Conditional exposure moments under two-stage Dirichlet assignment}\label{app:dirichlet-moments}

\begin{lemma}[Conditional exposure moments under two-stage Dirichlet assignment]
\label{lem:dirichlet-moments}
Fix a cluster $j$ with $N_j$ units and $M{+}1$ treatment levels indexed by
$m\in\{0,\ldots,M\}$. Consider the two-stage assignment:
\[
\Pi_j \sim \mathrm{Dirichlet}(\bar\alpha,\ldots,\bar\alpha),
\qquad
A_{j,i}\mid \Pi_j \;\overset{\text{iid}}{\sim}\; \mathrm{Categorical}(\Pi_j),
\]
so that $\Pr[A_{j,i}=m\mid \Pi_j]=\Pi_j[m]$ for each $i$.
Let
$p_{j[\ell]}(\A_j)=N_j^{-1}\sum_{r=1}^{N_j}\mathbf{1}\{A_{j,r}=\ell\}$ denote the realized
within-cluster share in arm $\ell$.

Then for any $\ell\in\{0,\ldots,M\}$ and any unit $i$,
\begin{align}
\E\!\left[\Pi_j[\ell]\mid A_{j,i}=m\right]
&=
\frac{\bar\alpha+\mathbf{1}\{\ell=m\}}{(M+1)\bar\alpha+1},
\label{eq:condPi-lem}\\[3pt]
\E\!\left[p_{j[\ell]}(\A_j)\mid A_{j,i}=m\right]
&=
\frac{1}{N_j}\mathbf{1}\{\ell=m\}
+\frac{N_j-1}{N_j}\cdot
\frac{\bar\alpha+\mathbf{1}\{\ell=m\}}{(M+1)\bar\alpha+1}.
\label{eq:condp-lem}
\end{align}
\end{lemma}

\begin{proof}
Write $\pi=(\pi[0],\ldots,\pi[M])$ for a generic point on the simplex
$\Delta^M=\{\pi:\pi[\ell]\ge 0,\ \sum_{\ell=0}^M \pi[\ell]=1\}$.
The Dirichlet$(\bar\alpha,\ldots,\bar\alpha)$ prior has density
\[
f(\pi)
=
\frac{\Gamma((M+1)\bar\alpha)}{\prod_{\ell=0}^M \Gamma(\bar\alpha)}
\prod_{\ell=0}^M \pi[\ell]^{\bar\alpha-1},
\qquad \pi\in\Delta^M.
\]
Under the Categorical likelihood, for a single observation $A_{j,i}=m$ we have
\[
\Pr[A_{j,i}=m\mid \Pi_j=\pi]=\pi[m].
\]
Therefore Bayes' rule gives the posterior density
\[
f(\pi\mid A_{j,i}=m)
\propto
\Pr[A_{j,i}=m\mid \Pi_j=\pi]\,f(\pi)
=
\pi[m]\cdot
\frac{\Gamma((M+1)\bar\alpha)}{\prod_{\ell=0}^M \Gamma(\bar\alpha)}
\prod_{\ell=0}^M \pi[\ell]^{\bar\alpha-1}.
\]
Absorbing constants into the normalizing factor, this becomes
\[
f(\pi\mid A_{j,i}=m)\ \propto\ 
\pi[m]^{\bar\alpha}\prod_{\ell\neq m}\pi[\ell]^{\bar\alpha-1},
\]
which is exactly the kernel of a Dirichlet distribution with parameters
\[
(\bar\alpha,\ldots,\bar\alpha,\bar\alpha+1,\bar\alpha,\ldots,\bar\alpha),
\]
i.e.,
\[
\Pi_j\mid(A_{j,i}=m)\sim
\mathrm{Dirichlet}\big(\bar\alpha+\mathbf{1}\{0=m\},\ldots,\bar\alpha+\mathbf{1}\{M=m\}\big).
\]
A standard property of the Dirichlet distribution is that if
$X\sim \mathrm{Dirichlet}(\alpha_0,\ldots,\alpha_M)$ and $\alpha_+\equiv\sum_{\ell=0}^M\alpha_\ell$,
then $\E[X[\ell]]=\alpha_\ell/\alpha_+$.
Applying this to the posterior above (whose total mass is $(M+1)\bar\alpha+1$) yields \eqref{eq:condPi-lem}.

For \eqref{eq:condp-lem}, by definition,
\[
p_{j[\ell]}(\A_j)=\frac{1}{N_j}\sum_{r=1}^{N_j}\mathbf{1}\{A_{j,r}=\ell\}
=\frac{1}{N_j}\mathbf{1}\{A_{j,i}=\ell\}
+\frac{1}{N_j}\sum_{r\neq i}\mathbf{1}\{A_{j,r}=\ell\}.
\]
Conditioning on $A_{j,i}=m$ makes the first term deterministic:
$\mathbf{1}\{A_{j,i}=\ell\}=\mathbf{1}\{\ell=m\}$.
For any $r\neq i$, use iterated expectations and conditional independence given $\Pi_j$:
\begin{align*}
\E\!\left[\mathbf{1}\{A_{j,r}=\ell\}\mid A_{j,i}=m\right]
&=
\E\!\left[
\E\!\left[\mathbf{1}\{A_{j,r}=\ell\}\mid \Pi_j,\,A_{j,i}=m\right]
\Bigm|\,
A_{j,i}=m
\right]\\
&=
\E\!\left[
\Pr[A_{j,r}=\ell\mid \Pi_j,\,A_{j,i}=m]
\Bigm|\,
A_{j,i}=m
\right]\\
&=
\E\!\left[
\Pr[A_{j,r}=\ell\mid \Pi_j]
\Bigm|\,
A_{j,i}=m
\right]
=
\E\!\left[\Pi_j[\ell]\mid A_{j,i}=m\right],
\end{align*}
where the third equality uses that $A_{j,r}\perp A_{j,i}\mid \Pi_j$.
Substituting into the decomposition of $p_{j[\ell]}(\A_j)$ gives
\[
\E\!\left[p_{j[\ell]}(\A_j)\mid A_{j,i}=m\right]
=
\frac{1}{N_j}\mathbf{1}\{\ell=m\}
+\frac{N_j-1}{N_j}\,\E\!\left[\Pi_j[\ell]\mid A_{j,i}=m\right].
\]
Plugging in \eqref{eq:condPi-lem} yields \eqref{eq:condp-lem}.
\end{proof}
}

\clearpage
\section{Additional simulation results}\label{app:table}

\rev{Appendix Table~\ref{table:homogeneous_sims_summary} reports the minimum-RMSE design (in terms of $\rho_T$) for each estimator across simulation scenarios.} \revii{The accompanying bias, RMSE, estimated standard error, and coverage in that table are evaluated separately at each estimator's selected grid value. Appendix Table~\ref{table:homogeneous_sims} reports selected results at three representative values of $\rho_T$ for each $(\rho_u,c)$ combination; the minimum-RMSE designs are selected over the full simulated grid. Across all 340 candidate designs and 340,000 production replications, every initial assignment was valid and both LM uniform-policy contrasts were estimable, so no redraw or alternative estimate was needed.}

\input{tables/homogeneous_sim_results_summary.tex}

\clearpage

\input{tables/homogeneous_sim_results.tex}

\clearpage
\rev{
\section{Additional application details}\label{app:application-details}

\subsection{Facebook video playback variants}\label{app:fb-variants}
The variants were organized as five matched \revii{pairs of low- and high-data configurations}: within each pair, the high-data configuration increases buffering intensity (and hence data usage) relative to the corresponding low-data configuration, while across pairs the remaining buffer parameters vary to cover multiple operating points within each data-usage level.
This paired structure keeps the arm count manageable while yielding a clear high-versus-low contrast and within-level variation, and it generates within-session variation in the share of high-data videos under \revii{user--video} randomization, which is central to our interference model.
Low-data configurations may be preferred for cost-constrained deployment or for users with limited connectivity.
}

%% file: tables/homogeneous_sim_results_summary.tex
\begin{table}[htbp]
\centering
\begin{adjustbox}{width=\textwidth}
\begin{tabular}{rrrrrrrrrrrrrr}
  \hline
  &&&&\multicolumn{2}{c}{Grid-minimizing $\rho_T(\bar \alpha)$} &\multicolumn{2}{c}{Mean absolute arm bias} & \multicolumn{2}{c}{Mean arm RMSE} & \multicolumn{2}{c}{Mean arm estimated SE} & \multicolumn{2}{c}{Mean arm coverage} \\
 $\rho_u$ & $c$ & $\Psi(1,0)$ & $\Psi(2,0)$ & LM & DM & LM & DM & LM & DM & LM & DM & LM & DM \\
0.000 & 0.000 & 2.500 & -2.500 & 0.999 & 0.250 & 0.000 & 0.000 & 0.035 & 0.034 & 0.035 & 0.034 & 0.956 & 0.950 \\ 
  0.100 & 0.000 & 2.500 & -2.500 & 0.999 & 0.001 & 0.003 & 0.001 & 0.090 & 0.037 & 0.088 & 0.036 & 0.941 & 0.942 \\ 
  0.300 & 0.000 & 2.500 & -2.500 & 0.999 & 0.001 & 0.005 & 0.001 & 0.167 & 0.041 & 0.163 & 0.042 & 0.939 & 0.956 \\ 
  0.500 & 0.000 & 2.500 & -2.500 & 0.990 & 0.001 & 0.005 & 0.001 & 0.248 & 0.048 & 0.249 & 0.049 & 0.948 & 0.953 \\ 
  0.800 & 0.000 & 2.500 & -2.500 & 0.999 & 0.001 & 0.005 & 0.004 & 0.486 & 0.078 & 0.489 & 0.078 & 0.950 & 0.952 \\ 
  0.000 & 0.100 & 2.600 & -2.750 & 0.999 & 0.999 & 0.001 & 0.001 & 0.035 & 0.035 & 0.035 & 0.035 & 0.942 & 0.943 \\ 
  0.100 & 0.100 & 2.600 & -2.750 & 0.999 & 0.990 & 0.004 & 0.004 & 0.088 & 0.087 & 0.088 & 0.088 & 0.946 & 0.947 \\ 
  0.300 & 0.100 & 2.600 & -2.750 & 0.999 & 0.625 & 0.008 & 0.061 & 0.163 & 0.146 & 0.164 & 0.132 & 0.944 & 0.914 \\ 
  0.500 & 0.100 & 2.600 & -2.750 & 0.999 & 0.001 & 0.005 & 0.170 & 0.250 & 0.179 & 0.246 & 0.049 & 0.948 & 0.250 \\ 
  0.800 & 0.100 & 2.600 & -2.750 & 0.999 & 0.001 & 0.014 & 0.172 & 0.489 & 0.192 & 0.489 & 0.078 & 0.941 & 0.431 \\ 
  0.000 & 0.500 & 3.000 & -3.750 & 0.999 & 0.999 & 0.001 & 0.001 & 0.035 & 0.035 & 0.035 & 0.035 & 0.938 & 0.937 \\ 
  0.100 & 0.500 & 3.000 & -3.750 & 0.999 & 0.999 & 0.002 & 0.001 & 0.087 & 0.087 & 0.088 & 0.088 & 0.948 & 0.948 \\ 
  0.300 & 0.500 & 3.000 & -3.750 & 0.999 & 0.952 & 0.003 & 0.046 & 0.168 & 0.166 & 0.163 & 0.161 & 0.942 & 0.941 \\ 
  0.500 & 0.500 & 3.000 & -3.750 & 0.990 & 0.990 & 0.002 & 0.007 & 0.251 & 0.247 & 0.248 & 0.244 & 0.944 & 0.940 \\ 
  0.800 & 0.500 & 3.000 & -3.750 & 0.999 & 0.769 & 0.011 & 0.195 & 0.502 & 0.477 & 0.489 & 0.429 & 0.941 & 0.917 \\ 
  0.000 & 1.000 & 3.500 & -5.000 & 0.999 & 0.999 & 0.001 & 0.002 & 0.035 & 0.035 & 0.034 & 0.035 & 0.952 & 0.950 \\ 
  0.100 & 1.000 & 3.500 & -5.000 & 0.999 & 0.999 & 0.001 & 0.001 & 0.089 & 0.089 & 0.088 & 0.088 & 0.941 & 0.940 \\ 
  0.300 & 1.000 & 3.500 & -5.000 & 0.990 & 0.990 & 0.004 & 0.013 & 0.168 & 0.166 & 0.166 & 0.165 & 0.944 & 0.948 \\ 
  0.500 & 1.000 & 3.500 & -5.000 & 0.990 & 0.990 & 0.005 & 0.022 & 0.246 & 0.245 & 0.250 & 0.246 & 0.946 & 0.948 \\ 
  0.800 & 1.000 & 3.500 & -5.000 & 0.990 & 0.990 & 0.011 & 0.006 & 0.490 & 0.483 & 0.493 & 0.486 & 0.950 & 0.949 \\ 
   \hline
\end{tabular}
\end{adjustbox}
\caption{Simulation results at each estimator's minimum-RMSE design on the simulated $\rho_T$ grid. For each estimator, mean absolute arm bias, RMSE, estimated SE, and coverage are evaluated at that estimator's selected grid value and are arithmetic means across the two treatment-versus-control arms.} 
\label{table:homogeneous_sims_summary}
\end{table}

%% file: tables/homogeneous_sim_results.tex
\begin{table}[htbp]
\centering
\begin{adjustbox}{totalheight=\textheight-6\baselineskip,width=\textwidth,keepaspectratio}
\begin{tabular}{rrrrrrrrrrrrrr}
  \hline
  &&&&&\multicolumn{2}{c}{Mean absolute arm bias} & \multicolumn{2}{c}{Mean arm RMSE} & \multicolumn{2}{c}{Mean arm estimated SE} & \multicolumn{2}{c}{Mean arm coverage} & \\
 $\rho_u$ & $c$ & $\rho_T$ & $\Psi(1,0)$ & $\Psi(2,0)$ & LM & DM & LM & DM & LM & DM & LM & DM & Initial valid rate \\
0.000 & 0.000 & 0.990 & 2.500 & -2.500 & 0.001 & 0.001 & 0.035 & 0.035 & 0.035 & 0.035 & 0.944 & 0.948 & 1.000 \\ 
  0.000 & 0.000 & 0.526 & 2.500 & -2.500 & 0.001 & 0.001 & 0.053 & 0.035 & 0.052 & 0.034 & 0.934 & 0.945 & 1.000 \\ 
  0.000 & 0.000 & 0.001 & 2.500 & -2.500 & 0.006 & 0.001 & 0.348 & 0.034 & 0.333 & 0.034 & 0.935 & 0.952 & 1.000 \\ 
  0.100 & 0.000 & 0.990 & 2.500 & -2.500 & 0.000 & 0.000 & 0.091 & 0.089 & 0.089 & 0.088 & 0.942 & 0.942 & 1.000 \\ 
  0.100 & 0.000 & 0.526 & 2.500 & -2.500 & 0.002 & 0.001 & 0.132 & 0.068 & 0.127 & 0.068 & 0.940 & 0.949 & 1.000 \\ 
  0.100 & 0.000 & 0.001 & 2.500 & -2.500 & 0.023 & 0.001 & 0.685 & 0.037 & 0.661 & 0.036 & 0.935 & 0.942 & 1.000 \\ 
  0.300 & 0.000 & 0.990 & 2.500 & -2.500 & 0.004 & 0.004 & 0.168 & 0.166 & 0.164 & 0.162 & 0.942 & 0.940 & 1.000 \\ 
  0.300 & 0.000 & 0.526 & 2.500 & -2.500 & 0.004 & 0.002 & 0.243 & 0.124 & 0.234 & 0.121 & 0.941 & 0.950 & 1.000 \\ 
  0.300 & 0.000 & 0.001 & 2.500 & -2.500 & 0.025 & 0.001 & 1.214 & 0.041 & 1.152 & 0.042 & 0.927 & 0.956 & 1.000 \\ 
  0.500 & 0.000 & 0.990 & 2.500 & -2.500 & 0.005 & 0.005 & 0.248 & 0.243 & 0.249 & 0.245 & 0.948 & 0.950 & 1.000 \\ 
  0.500 & 0.000 & 0.526 & 2.500 & -2.500 & 0.010 & 0.004 & 0.362 & 0.184 & 0.354 & 0.181 & 0.941 & 0.944 & 1.000 \\ 
  0.500 & 0.000 & 0.001 & 2.500 & -2.500 & 0.066 & 0.001 & 1.771 & 0.048 & 1.726 & 0.049 & 0.934 & 0.953 & 1.000 \\ 
  0.800 & 0.000 & 0.990 & 2.500 & -2.500 & 0.020 & 0.018 & 0.505 & 0.496 & 0.493 & 0.486 & 0.938 & 0.937 & 1.000 \\ 
  0.800 & 0.000 & 0.526 & 2.500 & -2.500 & 0.010 & 0.004 & 0.721 & 0.357 & 0.700 & 0.356 & 0.940 & 0.944 & 1.000 \\ 
  0.800 & 0.000 & 0.001 & 2.500 & -2.500 & 0.122 & 0.004 & 3.395 & 0.078 & 3.411 & 0.078 & 0.945 & 0.952 & 1.000 \\ 
  0.000 & 0.100 & 0.990 & 2.600 & -2.750 & 0.001 & 0.002 & 0.037 & 0.037 & 0.035 & 0.035 & 0.933 & 0.929 & 1.000 \\ 
  0.000 & 0.100 & 0.526 & 2.600 & -2.750 & 0.001 & 0.083 & 0.053 & 0.091 & 0.052 & 0.036 & 0.940 & 0.426 & 1.000 \\ 
  0.000 & 0.100 & 0.001 & 2.600 & -2.750 & 0.003 & 0.172 & 0.338 & 0.176 & 0.334 & 0.035 & 0.942 & 0.091 & 1.000 \\ 
  0.100 & 0.100 & 0.990 & 2.600 & -2.750 & 0.002 & 0.004 & 0.088 & 0.087 & 0.090 & 0.088 & 0.946 & 0.947 & 1.000 \\ 
  0.100 & 0.100 & 0.526 & 2.600 & -2.750 & 0.005 & 0.085 & 0.132 & 0.112 & 0.127 & 0.069 & 0.932 & 0.726 & 1.000 \\ 
  0.100 & 0.100 & 0.001 & 2.600 & -2.750 & 0.029 & 0.173 & 0.682 & 0.177 & 0.659 & 0.036 & 0.937 & 0.106 & 1.000 \\ 
  0.300 & 0.100 & 0.990 & 2.600 & -2.750 & 0.006 & 0.007 & 0.171 & 0.168 & 0.166 & 0.163 & 0.936 & 0.934 & 1.000 \\ 
  0.300 & 0.100 & 0.526 & 2.600 & -2.750 & 0.009 & 0.086 & 0.242 & 0.153 & 0.234 & 0.121 & 0.927 & 0.879 & 1.000 \\ 
  0.300 & 0.100 & 0.001 & 2.600 & -2.750 & 0.017 & 0.171 & 1.189 & 0.177 & 1.165 & 0.042 & 0.937 & 0.165 & 1.000 \\ 
  0.500 & 0.100 & 0.990 & 2.600 & -2.750 & 0.004 & 0.004 & 0.252 & 0.248 & 0.248 & 0.244 & 0.942 & 0.942 & 1.000 \\ 
  0.500 & 0.100 & 0.526 & 2.600 & -2.750 & 0.011 & 0.078 & 0.364 & 0.203 & 0.352 & 0.180 & 0.925 & 0.915 & 1.000 \\ 
  0.500 & 0.100 & 0.001 & 2.600 & -2.750 & 0.054 & 0.170 & 1.758 & 0.179 & 1.730 & 0.049 & 0.932 & 0.250 & 1.000 \\ 
  0.800 & 0.100 & 0.990 & 2.600 & -2.750 & 0.004 & 0.006 & 0.511 & 0.503 & 0.493 & 0.485 & 0.941 & 0.940 & 1.000 \\ 
  0.800 & 0.100 & 0.526 & 2.600 & -2.750 & 0.006 & 0.075 & 0.728 & 0.367 & 0.697 & 0.355 & 0.936 & 0.939 & 1.000 \\ 
  0.800 & 0.100 & 0.001 & 2.600 & -2.750 & 0.100 & 0.172 & 3.479 & 0.192 & 3.418 & 0.078 & 0.934 & 0.431 & 1.000 \\ 
  0.000 & 0.500 & 0.990 & 3.000 & -3.750 & 0.001 & 0.008 & 0.036 & 0.037 & 0.035 & 0.036 & 0.946 & 0.942 & 1.000 \\ 
  0.000 & 0.500 & 0.526 & 3.000 & -3.750 & 0.002 & 0.408 & 0.054 & 0.413 & 0.052 & 0.059 & 0.935 & 0.001 & 1.000 \\ 
  0.000 & 0.500 & 0.001 & 3.000 & -3.750 & 0.005 & 0.857 & 0.341 & 0.858 & 0.336 & 0.036 & 0.942 & 0.000 & 1.000 \\ 
  0.100 & 0.500 & 0.990 & 3.000 & -3.750 & 0.003 & 0.006 & 0.091 & 0.090 & 0.089 & 0.088 & 0.935 & 0.932 & 1.000 \\ 
  0.100 & 0.500 & 0.526 & 3.000 & -3.750 & 0.001 & 0.408 & 0.128 & 0.417 & 0.127 & 0.084 & 0.941 & 0.061 & 1.000 \\ 
  0.100 & 0.500 & 0.001 & 3.000 & -3.750 & 0.027 & 0.856 & 0.685 & 0.857 & 0.654 & 0.038 & 0.931 & 0.000 & 1.000 \\ 
  0.300 & 0.500 & 0.990 & 3.000 & -3.750 & 0.005 & 0.009 & 0.170 & 0.169 & 0.165 & 0.163 & 0.934 & 0.938 & 1.000 \\ 
  0.300 & 0.500 & 0.526 & 3.000 & -3.750 & 0.002 & 0.410 & 0.241 & 0.433 & 0.234 & 0.131 & 0.940 & 0.267 & 1.000 \\ 
  0.300 & 0.500 & 0.001 & 3.000 & -3.750 & 0.030 & 0.857 & 1.222 & 0.859 & 1.159 & 0.042 & 0.926 & 0.000 & 1.000 \\ 
  0.500 & 0.500 & 0.990 & 3.000 & -3.750 & 0.002 & 0.007 & 0.251 & 0.247 & 0.248 & 0.244 & 0.944 & 0.940 & 1.000 \\ 
  0.500 & 0.500 & 0.526 & 3.000 & -3.750 & 0.018 & 0.409 & 0.373 & 0.458 & 0.351 & 0.186 & 0.928 & 0.445 & 1.000 \\ 
  0.500 & 0.500 & 0.001 & 3.000 & -3.750 & 0.118 & 0.857 & 1.778 & 0.859 & 1.729 & 0.050 & 0.936 & 0.000 & 1.000 \\ 
  0.800 & 0.500 & 0.990 & 3.000 & -3.750 & 0.027 & 0.022 & 0.503 & 0.493 & 0.494 & 0.487 & 0.946 & 0.950 & 1.000 \\ 
  0.800 & 0.500 & 0.526 & 3.000 & -3.750 & 0.011 & 0.415 & 0.745 & 0.571 & 0.696 & 0.359 & 0.933 & 0.750 & 1.000 \\ 
  0.800 & 0.500 & 0.001 & 3.000 & -3.750 & 0.047 & 0.856 & 3.495 & 0.861 & 3.413 & 0.078 & 0.938 & 0.000 & 1.000 \\ 
  0.000 & 1.000 & 0.990 & 3.500 & -5.000 & 0.002 & 0.016 & 0.036 & 0.044 & 0.035 & 0.038 & 0.936 & 0.925 & 1.000 \\ 
  0.000 & 1.000 & 0.526 & 3.500 & -5.000 & 0.001 & 0.816 & 0.054 & 0.823 & 0.052 & 0.100 & 0.938 & 0.000 & 1.000 \\ 
  0.000 & 1.000 & 0.001 & 3.500 & -5.000 & 0.003 & 1.713 & 0.341 & 1.714 & 0.335 & 0.039 & 0.941 & 0.000 & 1.000 \\ 
  0.100 & 1.000 & 0.990 & 3.500 & -5.000 & 0.001 & 0.018 & 0.093 & 0.095 & 0.090 & 0.090 & 0.943 & 0.939 & 1.000 \\ 
  0.100 & 1.000 & 0.526 & 3.500 & -5.000 & 0.005 & 0.816 & 0.131 & 0.825 & 0.127 & 0.118 & 0.938 & 0.000 & 1.000 \\ 
  0.100 & 1.000 & 0.001 & 3.500 & -5.000 & 0.021 & 1.714 & 0.690 & 1.715 & 0.656 & 0.041 & 0.927 & 0.000 & 1.000 \\ 
  0.300 & 1.000 & 0.990 & 3.500 & -5.000 & 0.004 & 0.013 & 0.168 & 0.166 & 0.166 & 0.165 & 0.944 & 0.948 & 1.000 \\ 
  0.300 & 1.000 & 0.526 & 3.500 & -5.000 & 0.005 & 0.817 & 0.239 & 0.833 & 0.234 & 0.156 & 0.934 & 0.031 & 1.000 \\ 
  0.300 & 1.000 & 0.001 & 3.500 & -5.000 & 0.018 & 1.714 & 1.184 & 1.715 & 1.151 & 0.045 & 0.944 & 0.000 & 1.000 \\ 
  0.500 & 1.000 & 0.990 & 3.500 & -5.000 & 0.005 & 0.022 & 0.246 & 0.245 & 0.250 & 0.246 & 0.946 & 0.948 & 1.000 \\ 
  0.500 & 1.000 & 0.526 & 3.500 & -5.000 & 0.006 & 0.807 & 0.363 & 0.836 & 0.352 & 0.206 & 0.936 & 0.160 & 1.000 \\ 
  0.500 & 1.000 & 0.001 & 3.500 & -5.000 & 0.022 & 1.713 & 1.793 & 1.714 & 1.726 & 0.052 & 0.939 & 0.000 & 1.000 \\ 
  0.800 & 1.000 & 0.990 & 3.500 & -5.000 & 0.011 & 0.006 & 0.490 & 0.483 & 0.493 & 0.486 & 0.950 & 0.949 & 1.000 \\ 
  0.800 & 1.000 & 0.526 & 3.500 & -5.000 & 0.004 & 0.817 & 0.733 & 0.914 & 0.700 & 0.370 & 0.933 & 0.437 & 1.000 \\ 
  0.800 & 1.000 & 0.001 & 3.500 & -5.000 & 0.066 & 1.714 & 3.533 & 1.717 & 3.431 & 0.080 & 0.942 & 0.000 & 1.000 \\ 
   \hline
\end{tabular}
\end{adjustbox}
\caption{Selected simulation results at three representative values of $\rho_T$. Mean absolute arm bias, RMSE, estimated SE, and coverage are arithmetic means across the two treatment-versus-control arms; the last column reports the initial valid-assignment rate.} 
\label{table:homogeneous_sims}
\end{table}